\documentclass[pra,final,twocolumn,showpacs,aps]{revtex4-2}
\usepackage{eurosym}
\usepackage{multirow}
\usepackage{graphicx}
\usepackage{amssymb}
\usepackage{amsmath}
\usepackage{color}
\usepackage{xcolor}
\usepackage[table]{xcolor}
\usepackage{multirow}
\usepackage{afterpage}
\usepackage{hyperref}
\usepackage{float}
\usepackage{ulem}

\begin{document}
\title{Stable Quantum Vortices in Lee-Huang-Yang Dipolar Superfluids}
\author{S. Sabari$^{(a)}$, R. Radha$^{(b)}$, Lauro Tomio$^{(a)}$ and B. A.
Malomed$^{(c, d)}$}
\address{{$^{(a)}$Instituto de F\'\i sica Te\'{o}rica, Universidade Estadual Paulista,
01140-070 S\~{a}o Paulo, SP, Brazil.\\}
$^{(b)}$Centre for Nonlinear Science(CeNsc), Department of Physics, Government College
for Women(A), Kumbakonam 612001, India.\\
$^{(c)}$Department of Physical Electronics, School of Electrical Engineering,
Faculty of Engineering, and Center for
Light-Matter Interaction, Tel Aviv University, P.O.B. 39040, Ramat Aviv, 
Tel Aviv, Israel.\\
$^{(d)}$Instituto de Alta Investigaci\'{o}n, Universidad de Tarapac\'{a}, 
Casilla 7D, Arica, Chile.}
\date{\today }

\begin{abstract}
The nucleation and dynamics of vortices in the quasi-two-dimensional rotating 
dipolar Bose-Einstein condensate are explored by taking into account the 
Lee-Huang-Yang (LHY) correction to the mean-field (MF) theory. Assuming 
approximate cancellation of the MF interactions, we focus on the formation of 
a pure LHY superfluid. The effect of rotational frequency $\Omega $ is 
investigated numerically by determining the corresponding number of stable 
vortices in the superfluid, together with the respective energy per particle 
$E$ and chemical potential $\mu $. The LHY superfluid provides a deep minimum 
of $E$ and $\mu $, indicating that it is a remarkably robust state of quantum
matter. By fixing the LHY interaction strength, an exact single-vortex
critical frequency is found, along with the respective chemical potential. A
notable feature, observed when creating the LHY superfluid with fewer than five 
vortices, which is understood as being due to the superfluid's nonlinearity and 
trapping aspect ratio, is the 
large frequency ranges admitting the production of two and four vortices, as 
compared to the small frequency ranges to obtain one and three vortices. 
\end{abstract}

\maketitle


\section{Introduction}

\label{sec1} The experimental observation of dipolar Bose-Einstein
condensates (BECs) in ultracold gases of $^{52}$Cr~\cite 
{Koch:2008,Lahaye:2007,Griesmaier:2006}, $^{164}$Dy~\cite{Lu:2011,Youn:2010}
and $^{168}$Er~\cite{Aikawa:2012} magnetic atoms has been a giant step
forward in the studies of quantum matter~\cite{dbec1,dbec2}. BECs of
magnetic atoms exhibit long-range dipole-dipole interaction (DDI) between
atoms, depending on the mutual orientation of atomic magnetic moments \cite 
{2002Giovanazzi}, in addition to nonlinear effects induced by the $s-$wave
contact interactions between the atoms, including repulsion and attraction.
On the other hand, techniques based on the Feshbach resonance can
efficiently control the contact nonlinearity~\cite{Feshbach2010}, enabling
the control of the interplay between the nonlocal (DDI) and local (contact)
interactions for performing various manipulations with the dipolar BECs. In
particular, this setting was elaborated in Ref.~\cite{2019Kumar} to control
spatial separation of spin-coupled BECs. In this regard, it is relevant to
note that tunable long-range interatomic correlations can be maintained by
laser-induced DDIs in BECs \cite{2003ODell}, creating a \textit{roton} 
\textit{minimum} in the excitation spectrum, reminiscent of the well-known
properties of the strongly correlated superfluid represented by liquid
helium II \cite{1955Feynman,Noziers}. Rotons are also well known as
elementary excitations in dipolar BECs, see~\cite{Santos}-\cite{rotons-new}
and references therein.

It is well known that the mean-field (MF) cubic self-attractions leads to
the onset of the critical or supercritical collapse and eventual breakdown
of the condensates,~in the framework of the corresponding two- and
three-dimensional (2D and 3D) Gross-Pitaevskii (GP) equations, respectively
\cite{Berge1998,Sulem,2001Gammal,Fibich}. Therefore, it is relevant to
search for physically relevant mechanisms that can stabilize 2D and 3D
self-trapped modes, such as bright solitons, solitary vortices, and others~ 
\cite{2020Kartashov,2022-Malomed-Book}. In this context, beyond-MF effects,
which have been the subject of intensive experimental investigations~\cite 
{Altmeyer2007,Shin2008,Papp2008}, have shown a great potential. In this
vein, it was predicted \cite{Petrov2015,Petrov2016} that the full stability
of 2D and 3D self-trapped states in a binary BEC, in the form of \textit{ 
quantum droplets }(QDs), may be provided by the Lee-Huang-Yang (LHY)
corrections to the GP equations \cite{Lee1957,Lee1957a}, which represent the
effect of quantum fluctuations around the MF state. This result follows a
previous suggestion~\cite{2002Bulgac}, made in the context of the three-body
Efimov physics~\cite{1973Efimov}, which pointed out that the BEC system with
attractive two-body interactions could eventually be stabilized against the
collapse by a repulsive three-body term~\cite{2000Gammal}, leading to the
formation of self-bound boson droplets, termed \textit{boselets} in Ref.
\cite{2002Bulgac}, so that a trapping potential is not needed to hold the
particles together.

In terms of the corresponding GP formalism, the correction to the MF
approximation, as considered in Ref.~\cite{Petrov2015}, is represented by
quartic self-repulsion term, which is usually much weaker than the cubic MF
one. The LHY correction becomes significant in the special case when the
intraspecies self-repulsion and interspecies cross-attraction terms in the
corresponding system of non-linearly coupled GP equations (mutuality fitted
by means of the Feshbach resonance \cite{Feshbach}) almost cancel each
other~for the symmetric state, with equal wave functions of the two
components \cite{Petrov2015}. In particular, the \textquotedblleft LHY
superfluid" was predicted \cite{Jorgensen2018} and observed \cite{Skov2021},
in which the full cancellation of the MF fields makes the entire
nonlinearity in the GP system represented solely by the LHY terms. 
In agreement with the predictions, the LHY corrections have made it possible to
create robust QDs in binary BECs with the contact interactions \cite 
{Cabrera2018,Cheiney,Inguscio,Inguscio2,D'Errico}, as well as in dipolar
condensates dominated by the long-range magnetic DDIs (see theoretical
predictions in Refs.~\cite{Edler,Baillie} and experimental realizations in
Refs. \cite{Kadau:2016,Ferrier,Chomaz:2016}). 
In the latter case, the interplay between long-range interactions and the LHY 
effect also makes it
possible to realize supersolids in the dipolar quantum 
gases~\cite{Tanzi:2019,Bottcher:2019,Chomaz:2019}, see also review 
\cite{Frontiers}. The effects of quantum fluctuations may support supersolids 
in other settings as well~\cite{Guo:2019,Sabari:2017,Tamil:2019,
Tanzi:2019b,Bottcher:2021,Hertkorn:2021}.
The role of the LHY corrections has also been demonstrated in experimental
studies of the critical BEC temperature~\cite{Smith:2011}, quantum depletion~ 
\cite{Lopes:2017}, excitation spectrum~\cite{Lopes:2017b}, and thermodynamic
equation of state~\cite{Navon:2011}. It is relevant to mention that, when
investigating the universality of the contact interactions in the regime of
observable beyond-MF effects, no measurable contributions from three-body
interactions were found \cite{Wild:2012}.

Quantized vortices, first predicted in the framework of GP equations for
trapped Bose gases in Ref. \cite{1999Butts}, have been created in
experiments using rotating magnetic traps or applying laser stirring to BEC
above a certain critical rotational frequency. The observation of the
vortices provides compelling evidence for the occurrence of superfluidity in
BECs, sharing some features with liquid helium~\cite{1991Donnelly}. In
dipolar Bose gases, quantized vortices have been intensively investigated,
within the MF approximation, in the last decades up to now~\cite 
{Yi2006,Vardi2,Wilson2009,Baranov2008,Klawunn2008,Klawunn,
Abad2009,Abad2010,Malet2011,Kishor2012,Wilson2012,Raymond1,Raymond2,Sabari:2018,
Sabari2024,Lauro2024,Sabari2017}. 
The structure and stability of vortices in dipolar BECs are strongly
affected by the anisotropy of the DDIs and their strengths, with respect to
the contact interactions~\cite{Yi2006,Wilson2009,Abad2009,Raymond3,Sabari2025}. Static
hydrodynamic solutions and their stability have been also explored for the
rotating dipolar BECs in the Thomas-Fermi (TF) limit~\cite{2009Bijnen}. In
this context, the structure of dipolar BECs was studied in the quasi-2D
scenario with arbitrary orientations of the dipoles~\cite{Vardi,Malet2011}.
The second-order-like phase transition of straight and helical vortex lines,
occurring due to the effect of the dipoles' orientation, has also been
reported \cite{Klawunn2008}. DDI strongly affects the number, structure, and
stability of vortices. Further, the critical rotational frequency for the
nucleation of vortices decreases with the increase of the DDI strength,
while the contact interactions enhance the vortex stability~\cite{Kishor2012} . 
The stabilization of 2D and 3D self-trapped (localized) QDs with embedded
vorticity due to the action of the LHY effect in the binary BECs with
contact interactions were predicted, respectively, in Refs.~\cite 
{2D-LHY-vort} and \cite{Leticia}. On the other hand, the QD solutions for
axially symmetric vortex states in the single-component dipolar BEC were
found to be fully unstable~\cite{Macri}.

Although experiments have been carried out with $^{164}$Dy and $^{168}$Er
dipolar condensates by taking into account the effects of quantum
fluctuations, quantized vortices in LHY-dominated superfluids have not yet
been explored in detail. The recent experimental observation of quantized
vortices in a dipolar condensate of lanthanide atoms~\cite 
{Klaus2022,2023Bland} motivates further theoretical studies
and new experiments with single- and multi-vortex states in one-component
and binary rotating condensates~\cite{2019Kumar} in the LHY-dominated
regime. In light of the above, the main objective of the present paper
is to report new results, based on systematic simulations, for the
nucleation and dynamics of vortices in a one-component dipolar BEC in the
LHY-dominated superfluid, featuring the interplay between dipolar and
contact two-body interactions. 
The independent tunability of binary interactions and dipole-dipole 
interactions ensures that one may end up with a fluid completely dominated 
by quantum fluctuations alone.

The paper is organized as follows. In the next section, we formulate the
single-component quasi-2D model including the DDI and LHY effect, basically
following Ref.~\cite{2019Kumar}. Systematic results, with the corresponding
discussion, are reported in Section III. 
Section IV summarizes the paper with our main conclusions and perspectives.

\section{The GP system with the DDI and LHY terms}

\begin{figure}[th]
\centerline{
\includegraphics[width=0.45\textwidth]{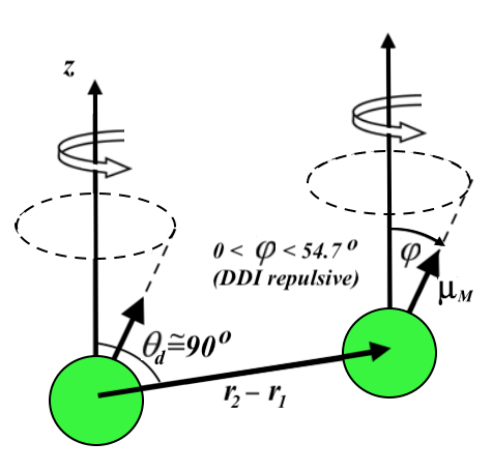}}
\caption{The representation of DDIs in the dipolar quasi-2D BEC in the
coordinate space. Angle between vector $\mathbf{r}_{2}\mathbf{-r}_{1}$ and
the $z$ axis is $\protect\theta _{d}\approx 90^{\mathrm{o}}$. The variation
of angle $\protect\varphi $ between the magnetic dipoles and the $z$ axis
alters the strength and sign of DDI, from repulsive to attractive. 
The circular arrows indicate rapid rotation of the magnetic moments.}
\label{fig01}
\end{figure}

{We consider a dipolar atomic BEC of $N$ atoms with mass $m$ and magnetic
dipole moment $\mu _{M}$. The system is confined in a quasi-2D
pancake-shaped harmonic-oscillator trap, } 
\begin{equation}
{V(\mathbf{r})\equiv \frac{m\omega_{\perp}^{2}}{2} \left(
x^{2}+y^{2}+\lambda ^{2}z^{2}\right) ,}  \label{V}
\end{equation} 
with a fixed large aspect ratio, $\lambda \equiv \omega _{z}/\omega _{\perp
}=20$. In the rotating reference frame, the dynamics of the system is
governed by the time-dependent GP equation~for the wavefunction $\phi \equiv
\phi ({\mathbf{r}},t)$, which is subject to the unitary normalization 
\cite{Lima2011,Blakie2016,Wachtler2016}:
\begin{eqnarray}
\mathrm{i}\hbar \frac{\partial \phi }{\partial t} &=&\Big[-\frac{\hbar ^{2}}{ 
2m}\nabla ^{2}+V(\mathbf{r})-\Omega L_{z}+g_{0}\left\vert \phi \right\vert
^{2}+g_{_{\mathrm{LHY}}}\left\vert \phi \right\vert ^{3}  \notag \\
&+&\int U_{\mathrm{dd}}({\mathbf{r}}-{\mathbf{r}}^{\prime })\left\vert \phi
^{\prime }\right\vert ^{2}d^{3}{\mathbf{r}}^{\prime }\Big]\phi ,
\label{eqn:3dgpe}
\end{eqnarray} 
where $L_{z}\equiv -\mathrm{i}\hbar \left( x\partial _{y}-y\partial
_{x}\right) $ is the $z$-component of the angular momentum operator, with $ 
\Omega L_{z}$ implying the rotation about the $z$-axis with angular
frequency $\Omega $. The coefficient $g_{0}$ of the cubic MF term is given
in terms of the two-body $s-$wave scattering length $a_{s}$, as $g_{0}={4\pi
\hbar ^{2}a_{s}N}/{m}$. The DDI is represented by the integral term in Eq.~ 
\eqref{eqn:3dgpe}, which refers to the quasi-2D geometry illustrated in Fig.~ 
\ref{fig01} \cite{2002Giovanazzi},
\begin{equation}
U_{\mathrm{dd}}(\mathbf{r-r^{\prime }})\equiv \displaystyle\frac{\mu _{0}{ 
\mu _{M}^{2}}N}{8\pi }\frac{\left( 3\cos ^{2}\varphi -1\right) }{|\mathbf{ 
r-r^{\prime }}|^{3}}.
\end{equation} 
The strength of the quartic nonlinear LHY term for the dipolar BEC is
\begin{equation}
g_{\mathrm{LHY}}=\frac{64\hbar ^{2}\sqrt{{\pi (a_{s}N)^{5}}}}{m}\left( \frac{ 
2}{3}+\varepsilon _{\mathrm{dd}}^{2}\right) ,  \label{glhy}
\end{equation} 
where $\varepsilon _{\mathrm{dd}}\equiv a_{\mathrm{dd}}/a_{s}$ is the ratio
of the effective DDI scattering length,
\begin{equation}
a_{\mathrm{dd}}\equiv\frac{m\mu_{0}\mu_{M}^{2}}
{12\pi \hbar ^{2}},\label{a_dd}
\end{equation} 
and the $s-$wave two-body scattering length $a_{s}$~\cite{dbec1}.

\subsection{DDI in the quasi-2D BEC}

To define the DDI, we consider a pair of atoms placed at positions $\mathbf{r 
}$ and $\mathbf{r^{\prime }}$, with the respective vector $\mathbf{ 
r-r^{\prime }}$ making angle $\theta _{d}$ with the $z$ axis, so that $ 
\theta _{d}\approx 90^{\circ }$ for the oblate (pancake-shaped)
configuration. As shown in Fig.~\ref{fig01}, the atoms have their magnetic
moments aligned, by a polarizing magnetic field, in the same direction,
making angle $\varphi $ with the $z$ axis. Similar to Refs.~\cite 
{2002Giovanazzi,2019Kumar}, in the present case, the tunability of the DDI
is provided by time-dependent polarizing fields, which set the magnetic
moments in rapid rotation around the $z$ axis. With the magnetic fields
being a combination of the static (dc) component directed along the $z$ axis
and the rapidly rotating (ac) one in the $(x,y)\ $plane, one may neglect the
motion of atoms driven by the ac field. With the time averaging performed
over the rotation period, the corresponding 3D averaged DDI can be written
in the scaled form as~\cite{2002Giovanazzi}
\begin{equation}
\hspace{-0.2cm}\left\langle V_{\mathrm{3D}}^{(d)}({\mathbf{r}-\mathbf{ 
r^{\prime }}})\right\rangle ={\mu _{0}\mu _{M}^{2}}\left( \frac{{3\cos
^{2}\varphi }-1}{2}\right) \frac{1}{\left\vert {\mathbf{r}-\mathbf{r^{\prime
}}}\right\vert ^{3}},  \label{DDI}
\end{equation} 
where $\mu _{0}$ is the free-space permeability, with the magnetic moment $ 
\mu _{M}$ measured in units of the Bohr magneton $\mu _{B}$. As said above,
for the oblate configuration adopted here, with the condensate distributed
in the $(x,y)$ plane, we have $\theta _{d}\approx 90^{\circ }$, hence $ 
\varphi $ remains an appropriate angular coordinate for tuning the DDI
strength, making it repulsive in the interval of $0\leq \varphi <\varphi _{ 
\mathrm{magic}}$, and attractive for $\varphi _{\mathrm{\ magic}}$ ${ 
<\varphi \leq 90}$ $^{\circ }$, where $\varphi _{\mathrm{magic}}\approx
54.7^{\circ }$ is the so-called \textquotedblleft magic angle" \cite{magic}.
DDI vanishes at $\varphi =\varphi _{\mathrm{magic}}$, irrespective of the
magnitude of the magnetic moment.

Equations~\eqref{eqn:3dgpe} and \eqref{glhy} demonstrate that the
short-range binary interaction, long-range DDI, and LHY correction are
interdependent, which can impede the search for stable configurations, -
in particular, vortices. Therefore, it may be helpful to develop a model
exhibiting three independent interactions, which do not impact on each
other. To this end, it is desirable to have the $g_{\mathrm{LHY}}$
coefficient which is affected not by the DDI orientation, but solely by the
corresponding DDI scattering length (\ref{a_dd}). Therefore, a simple
strategy to increase the impact of the LHY contribution is to assume a
scenario in which one can adjust the contact and dipolar interactions, so
that they effectively cancel out each other. To this aim, starting with a
repulsive contact interaction, the long-range DDI should be tuned by
adjusting the orientation angle $\varphi $ in Fig. \ref{fig01}, making it
attractive. Provided that the strength of the repulsive contact interaction
is adjusted to stay within the possible range of the attractive DDI
strength, it may be possible to identify the setting in which the LHY effect 
dominates the condensate. Similarly, with repulsive DDI, one can
manipulate the contact interaction, tuning it to the attraction sign, again
designing the LHY-controlled setting, with the dipole-dipole and contact MF
interactions balancing each other.

Appropriate dimensionless parameters can be defined in terms of a reference
frequency, which we set as $\omega _{\perp }=\omega _{x}=\omega _{y}$, see
Eq. (\ref{V}), and the corresponding oscillator length, $\ell _{\perp }= 
\sqrt{\hbar /(m\omega _{\perp })}$. To cast Eq. (\ref{eqn:3dgpe}) in the
normalized form, we first rescale the time and coordinate variables as $\bar{ 
t}\equiv \omega _{\perp }t$, $\bar{\mathbf{r}}\equiv \mathbf{r}$.
Accordingly, we define $\bar{a_{s}}=a_{s}/\ell _{\perp }$, $\bar{a}_{\mathrm{ 
dd}}=a_{\mathrm{dd}}/\ell _{\perp }$, with the wavefunction and system's
energy rescaled as $\bar{\phi}=\ell _{\perp }^{3/2}\phi $ and $\bar{E}=\hbar
\omega _{\perp }E$. Next, we drop the overhead bars, assuming that all the
variables and parameters are dimensionless. This transformation casts the GP
equation \eqref{eqn:3dgpe} into the normalized form, with $\hbar =m=1$ and
redefined coefficients.

In our approach, we assume a strong harmonic trap confinement in the
$z$-direction, allowing the 3D ground-state wavefunction be expressed 
by the ansatz $\phi(\mathbf{r},t)\equiv {(\lambda /\pi)^{1/4}}
\exp \left( {-\lambda z^{2}}/{2}\right) \psi (\boldsymbol{\rho },t),$
where $\psi \equiv \psi (\boldsymbol{\rho },t)$ is the effective 2D MF 
wavefunction in polar coordinates, where $\boldsymbol{\rho }\equiv (x,y)
= (\rho \cos\theta,\rho \sin\theta)$.
The factorized ansatz is substituted in 
Eq.~\eqref{eqn:3dgpe}, which is then multiplied by $\phi (z)\equiv $ ${(\lambda
/\pi )^{1/4}}\exp \left( {-\lambda z^{2}}/{2}\right) $ and integrated over $ 
z $~\cite{2017Kumar}. The corresponding procedure for the 2D reduction of
the DDI term is elaborated in detail in Refs.~\cite{Wilson2012,2019Kumar},
in which the effective 2D DDI was derived by applying the convolution
theorem and using the 2D Fourier transform for the product 
of the DDI kernel and density,
as $\int d{{\ \boldsymbol{\rho }}^{\prime }}V^{(d)}({{ 
\boldsymbol{\rho }}-{\boldsymbol{\ \rho }}^{\prime }})|\psi ({\boldsymbol{ 
\rho ^{\prime }}})|^{2}=\mathcal{F}_{\mathrm{2D}}^{-1}[\widetilde{V} 
^{(d)}(k_{x},k_{y})\widetilde{n}(k_{x},k_{y})]$, where $\mathcal{F}_{\mathrm{ 
2D}}^{-1}$\ is the operator of the inverse Fourier transform.
As a result, the DDI kernel can be expressed as the combination of two
terms in the 2D momentum space, $\mathbf{k_{\rho }}=(k_{x},k_{y})=(k_{\rho
}\cos \theta _{k},k_{\rho }\sin \theta _{k})$, for a fixed orientation $ 
\varphi $ of the magnetic dipoles (Fig. \ref{fig01}), accounting for the
interactions perpendicular and parallel to the direction of the dipole
moments:
{\small\begin{eqnarray}
\widetilde{V}_{\perp }^{(d)}(k_{x},k_{y}) &=&2-3\sqrt{\frac{\pi }{2\lambda }} 
k_{\rho }\exp \left( \frac{k_{\rho }^{2}}{2\lambda }\right) \mathrm{erfc} 
\left( \frac{k_{\rho }}{\sqrt{2\lambda }}\right) ,  \notag \\
\widetilde{V}_{\parallel }^{(d)}(k_{x},k_{y}) &=&-1+3\frac{k_{x}^{2}}{ 
k_{\rho }}\sqrt{\frac{\pi }{2\lambda }}\exp \left( \frac{k_{\rho }^{2}}{ 
2\lambda }\right) \mathrm{erfc}\left( \frac{k_{\rho }}{\sqrt{2\lambda }} 
\right) ,  \label{DDI-perp-par}
\end{eqnarray} 
}where erfc($\xi $) is the standard complementary error function. With all
directions being possible in the $\left( x,y\right) $ plane, the angular
averaging amounts to $k_{x}^{2}=k_{\rho }^{2}\cos ^{2}\theta _{k}\rightarrow
k_{\rho }^{2}/2$, so that the effective total 2D DDI kernel can be written
in the momentum space as a function of $k_{\rho }$ and $\varphi $, \textit{ 
viz}, $V_{\varphi }(k_{\rho })\equiv\widetilde{V}^{(d)}(k_{x},k_{y})$:
{\small\begin{eqnarray}
V_{\varphi }(k_{\rho })&=&
\cos^{2}\!\varphi \widetilde{V}_{\perp }^{(d)}(k_{x},k_{y})+\sin ^{2}\!\varphi
\widetilde{V}_{\parallel }^{(d)}(k_{x},k_{y})  \label{DDI-2D} \\
&=&\frac{3\cos ^{2}\!\varphi -1}{2}\left[ 2-3\sqrt{\frac{\pi }{2\lambda }} 
\;k_{\rho }\mathrm{e}^{\frac{k_{\rho }^{2}}{2\lambda }}\mathrm{erfc}\left(
\frac{k_{\rho }}{\sqrt{2\lambda }}\right) \right] .  \notag
\end{eqnarray}
}Thus, we have the effective GP equation given by
\begin{eqnarray}
\mathrm{i}\frac{\partial \psi }{\partial t} &=&\biggr[-\frac{\nabla _{\rho
}^{2}}{2}+\frac{\rho ^{2}}{2}-\Omega L_{z}+g|\psi |^{2}+\eta \left\vert \psi
\right\vert ^{3}\biggr]\psi  \label{equ:gpe2d} \\
&+& g_{dd}(\varphi )\int \frac{d^{2}\mathbf{k}_{\rho }}{(2\pi )^{2}} 
e^{-i\mathbf{k}_{\rho }.\boldsymbol{\rho }}\widetilde{n}(\mathbf{k}_{\rho
},t)h_{\mathrm{2D}}\biggr(\frac{k_{\rho }}{\sqrt{2\lambda }}\biggr)\psi,
\notag
\end{eqnarray} 
where $g_{dd}(\varphi
)\equiv (g_{dd}/2){(3\cos ^{2}\!\varphi -1)}$. Here, the 
coefficient $\eta $, which is defined below by Eq. (\ref{eta}), corresponds 
to the reduced LHY dimensionless coefficient $g_{\mathrm{LHY}}$ given by Eq.  
\eqref{glhy} for the dipolar system. The other factors in Eq. (\ref 
{equ:gpe2d}) are~\cite{2012Muruganandam}
\begin{eqnarray}
\widetilde{n}(\mathbf{k}_{\rho },t) &=&\int d^{2}\boldsymbol{\rho }e^{i 
\mathbf{k}_{\rho }.\boldsymbol{\rho }}|\psi |^{2},  \notag \\
h_{\mathrm{2D}}(\xi )\Big|_{\xi \equiv \frac{k_{\rho }}{\sqrt{2\lambda }}}
&=&\left[ 2-3\sqrt{\pi }\xi \exp (\xi ^{2})\text{erfc}(\xi )\right] .
\label{h2d}
\end{eqnarray} 
The 2D GP equation (\ref{equ:gpe2d}) is elaborated for quasi-2D BEC in the
next subsection. Further, in the framework of the quasi-2D approximation,
the 3D two-body contact-interaction constant, $g_{0}$, is replaced in Eq. ( 
\ref{equ:gpe2d}) by the scaled one, $g\equiv \sqrt{8\pi \lambda }{Na_{s}}/{ 
\ell _{\perp }}$. The\ rescaling transforms the original DDI coefficient ${ 
\mu _{0}{\mu _{M}}^{2}}/(4\pi )$ into the {dimensionless one,
${g}_{\mathrm{dd}}={\sqrt{8\pi \lambda }Na_{\mathrm{dd}}/\ell _{\perp }}$, } 
were ${a_{\mathrm{dd}}}$ is the DDI scattering length (\ref{a_dd}) \cite 
{2017jpco}, One should note that the aspect ratio 
$\lambda$ also appears explicitly in the integrated 2D expression for $ 
V^{(d)}$~\cite{2019Kumar}. Considering the DDI tunability through the
orientation angle $\varphi $ of the magnetic moments, the DDI coefficient is
written below as a function of $\varphi $, ${g}_{\mathrm{dd}}(\varphi )$.
For the length rescaling, an alternative is to use the Bohr radius $a_{0}$ ($ 
=5.29\times 10^{-11}$m), taking into account that $\ell _{\perp }=10^{-6}$m$ 
\approx 1.89\times 10^{4}a_{0}$.

The total energy corresponding to Eq.~\eqref{equ:gpe2d} can also be written 
as~\cite{2020Brito}
\begin{eqnarray}
E &=&\int \!\!dxdy\left\{ \frac{1}{2}\left[ \left\vert \mathrm{i} 
\partial_x\psi-\Omega y\psi \right\vert ^{2}+\left\vert
\mathrm{i}
\partial_y
+\Omega x\psi \right\vert ^{2} 
\right] \right.  \notag \\
&+&\left. \frac{(1-{\Omega }^{2})(x^{2}+y^{2})}{2}|\psi |^{2}+\frac{g}{2} 
|\psi |^{4}+\frac{2}{5}\eta |\psi |^{5}\right\}  \label{En} \\
&+&\frac{g_{dd}(\varphi )}{2}\int \frac{d^{2}\mathbf{k}_{\rho }}{ 
(2\pi )^{2}}e^{-i\mathbf{k}_{\rho }.\boldsymbol{\rho }}\widetilde{n}(\mathbf{ 
k}_{\rho },t)h_{\mathrm{2D}}\biggr(\frac{k_{\rho }}{\sqrt{2\lambda }}\biggr) 
|\psi |^{2}\Bigg\}.  \notag
\end{eqnarray} 
In the limit of extremely tight confinement in the transverse direction, the
effective nonlinearity in the LHY-amended GPE is different from the quartic
term~\cite{Petrov2016}. Nevertheless, one may use the 2D equations with the
quartic term, such as Eq. (\ref{equ:gpe2d}), in relevant experimental
settings with moderately tight confinement \cite{Viskol}.

\subsection{The LHY correction for the quasi-2D BEC}

The LHY corrections can be taken into account in the usual MF 3D formalism
similar to how it was done in Ref. \cite{Jorgensen2018}, where the
correction to the ground-state energy density of a homogeneous weakly
repulsive Bose gas with contact interactions, is 
\begin{eqnarray}
\mathcal{E}_{\mathrm{LHY}}&=&\frac{\hbar ^{2}}{15m}{256\sqrt{\pi (|a_{s}|n)^{5} 
}}\\ 
&=&\frac{\hbar \omega _{\perp }}{\ell _{\perp }^{3}}\frac{256\sqrt{\pi }}{ 
15}{{\left( N\frac{|a_{s}|}{\ell _{\perp }}|\psi (x,y)|^{2}\sqrt{\frac{ 
\lambda }{\pi }}e^{-\lambda z^{2}}\right) ^{5/2}}}.\nonumber
\end{eqnarray} 
The corresponding LHY corrections are given by
\begin{equation}
H_{\mathrm{LHY}}\Psi =\left( \frac{128\sqrt{\pi }}{3}\sqrt{\frac{N|a_{s}|^{5} 
}{\ell _{\perp }^{5}}}\right) |\psi (x,y)|^{3}|\phi (z)|^{3}\Psi .
\label{H_LHY}
\end{equation} 
Multiplying this term by the complex conjugate of $\phi (z)$, and
integrating out the $z-$component, we obtain the following expression for
the scaled LHY correction term,
\begin{equation}
H_{\mathrm{LHY}}\psi =\frac{128}{3}\sqrt{\frac{2N^{3}}{5}}\left( \frac{ 
\lambda ^{3}}{\pi }\right) ^{1/4}\left( {\frac{|a_{s}|}{\ell _{\perp }}} 
\right) ^{5/2}|\psi |^{3}\psi ,
\end{equation} 
which provides the scaled coefficient corresponding to coefficient $g_{ 
\mathrm{LHY}}$ in Eq. \eqref{glhy}, in the absence of DDI. In the case of GP
equation for the dipolar gas, the coefficient $g_{\mathrm{LHY}}$ needs to be
multiplied by an additional factor, which depends on the DDI strength, as
shown in Refs.~\cite{2006Schutzhold,Lima2011,Blakie2016,Wachtler2016,2016Schmitt}, thus
giving rise to the interdependence between the LHY correction, contact
interactions, and DDI. Including the DDI, we thus derive the effective
coefficient for the scaled quasi-2D LHY correction as
\begin{equation}
\eta =\frac{64\sqrt{2}}{\sqrt{5\sqrt{\pi }}}\left( {\sqrt{\lambda }N} 
\right) ^{\frac{3}{2}}\left( \frac{|a_{s}|}{\ell _{\perp }}\right) ^{\frac{5}{2}}
\left(\frac{2}{3}+\varepsilon _{\mathrm{dd}}^{2}\right) .  \label{eta}
\end{equation} 
It appears in Eq. \eqref{glhy}, for a quasi-2D dipolar condensate with
aspect-ratio parameter $\lambda $, that was defined above in Eq. (\ref{V}).

\subsection{The pure LHY nonlinearity in the quasi-2D GP equation}

In the above derivation of Eq.\eqref{eta} it is adopted that the dipoles are
aligned in the same direction (Fig. \ref{fig01}), implying that the DDI is
maximized at $\varphi =0^{\circ }$ \cite{Lima2011}. As also seen in Fig.~\ref 
{fig01}, the effective DDI strength can be tuned 
by varying $\varphi $ from $0$ (maximum positive) to $\pi/2$ (maximum 
negative), being zero for 
$\varphi =\varphi _{\mathrm{magic}}$. To take advantage of the
tunability of DDI ~\cite{2002Giovanazzi}, we keep $\varphi $ as an
adjustable parameter.
For this purpose, the LHY parameter \eqref{eta} is redefined, as 
$\eta\equiv\eta(\varphi)$, with 
\begin{eqnarray}
\varepsilon_{\rm{dd}}\equiv\varepsilon_{\mathrm{dd}}(\varphi )\equiv \frac{a_{ 
\mathrm{dd}}}{a_{s}}\left( \frac{3\cos ^{2}\varphi -1}{2}\right) .
\label{epsphi}
\end{eqnarray}  
Thus, the effective LHY coefficient is affected by two adjustable 
parameters: $a_{s}$ and the angle $\varphi$. 
Independently of the signs of the contact and dipolar interactions, the
LHY contribution is positive.
The possibility to enhance the LHY contribution as compared with the
main nonlinear terms, is to make use of both control parameters, so that one
of the interactions is attractive, with the other being repulsive.
The ideal condition for a pure LHY superfluid, is the overall 
total cancellation of the MF and dipolar terms. In the framework of a
more realistic experimental setup, one can consider the 
situation in which we have an effective reduction of the two main
nonlinear terms such that the fluid's nonlinearity is dominated by the 
quantum fluctuations.
In the following  section, we consider the ideal condition
in which we can effectively remove the MF terms, \textit{viz}., DDI and 
contact ones, in Eqs. \eqref{equ:gpe2d} and \eqref{En}.
The result is the following stationary 2D GP equation for the quantum superfluid (QS)
with chemical potential $\mu _{\mathrm{QSF}}$, and the respective energy per
particle:
{\small
\begin{eqnarray}
\hspace{-0.5cm}\mu _{QS}\psi &=&\biggr[-\frac{1}{2}\left( 
{\partial_x^{2}}+
{\partial_y^{2}}\right) +\frac{ 
x^{2}+y^{2}}{2}-\Omega L_{z}+\eta (\varphi )\left\vert \psi \right\vert ^{3} 
\biggr]\psi ,  \notag \\
{E_{\mathrm{QS}}}&=&\int \!\!dxdy\Bigg\{\frac{1}{2}\left[ \left\vert
\mathrm{i}{\partial_x}-\Omega y\psi \right\vert
^{2}+\left\vert \mathrm{i}{\partial_y}+\Omega x\psi
\right\vert ^{2}\right]  \label{EnLHY} \\
&+&\frac{(1-{\Omega }^{2})(x^{2}+y^{2})}{2}|\psi |^{2}+\frac{2}{5}\eta
(\varphi )|\psi |^{5}\Bigg\}.  \notag
\end{eqnarray}
This possibility to effectively cancel the local and nonlocal MF nonlinear terms is
somewhat similarity to that proposed in Ref.~\cite{Jorgensen2018},
where a binary Bose mixture without DDI and rotation was considered,
therefore our studies of on rotational properties of LHY superfluid,
following below in section~\ref{secIIIC}, may be accordingly applied to the
setting addressed in Ref.~\cite{Jorgensen2018}. In addition, it is relevant
to mention that, for the 2D BEC trapped in the harmonic-oscillator
potential, the critical rotational frequency $\Omega _{c}$ for a single
vortex production has been analytically determined in Ref.~\cite{2001Fetter}, 
assuming the strong cubic nonlinearity obeying the Thomas-Fermi limit. 
In the present notation, the corresponding expression is
\begin{equation}
\Omega _{c}\approx \frac{5}{2}\ln \left( 0.67\sqrt{2\mu }\right) .
\label{critic}
\end{equation} 
In the case of more general nonlinear interactions, such as the present
case, which includes the DDI and LHY terms, the result is expected to be
similar, as the essential differences originating from the nonlinear
interactions, may be absorbed by the change of the chemical potential.

\section{Numerical results and discussion}

\label{secIII} The main objective of this work is to explore the nucleation
and dynamics of vortices in the BEC dominated by the LHY nonlinearity,
assuming the effective mutual cancellation of the MF contact interactions
and DDIs. In subsection A, the results are produced in the absence of the
LHY term, aiming to verify the rotation effect in the usual situation, where
the MF nonlinearity is taken into account. These results, which are
consistent with previous investigations, serve as a benchmark for the
comparison with the cases where the LHY term is the dominant one. {In that
subsection, the explicit dependence on the orientation angle $\varphi $ is
dropped in the DDI and LHY parameters, assuming that $\eta \equiv \eta
(\varphi )$ is given by Eq.~\eqref{epsphi} and ${g}_{\mathrm{dd}}\equiv g_{ 
\mathrm{dd}}(\varphi )$. Unless explicitly stated otherwise, this implies $ 
\varphi =0^{\circ }$. In particular, the pure LHY superfluid considered in
subsection C relies on the assumption of the effective cancellation of the
MF contact and DDI terms, making it relevant to keep the $\varphi $
dependence of the coefficients $\eta $ and $g_{dd}$}.

\subsection{Vortex modes in the absence of the LHY term}

\label{secIIIA} The numerical procedure to solve the 2D GP equation starts
with the split-step Crank-Nicholson method. Its advantage is that the
non-derivative nonlinear and linear terms can be treated very accurately, to
secure the precision and robustness of the numerical scheme. 
In the present case, the usual split-step Crank-Nicholson
method was combined with the fast Fourier transform approach, following
the prescription given in Refs.~\cite{Sabari2025,2017Kumar,2017jpco}. 
Here, we have used a numerical code with 512 $\times$ 512 grid size,
with spatial step sizes $\Delta x=$ $\Delta y=0.04$. 
The norm of the wave function is restored to be 1 at each time step 
$\Delta t=0.01$.

\begin{figure}[th]
\begin{center}
\includegraphics[width=8.5cm]{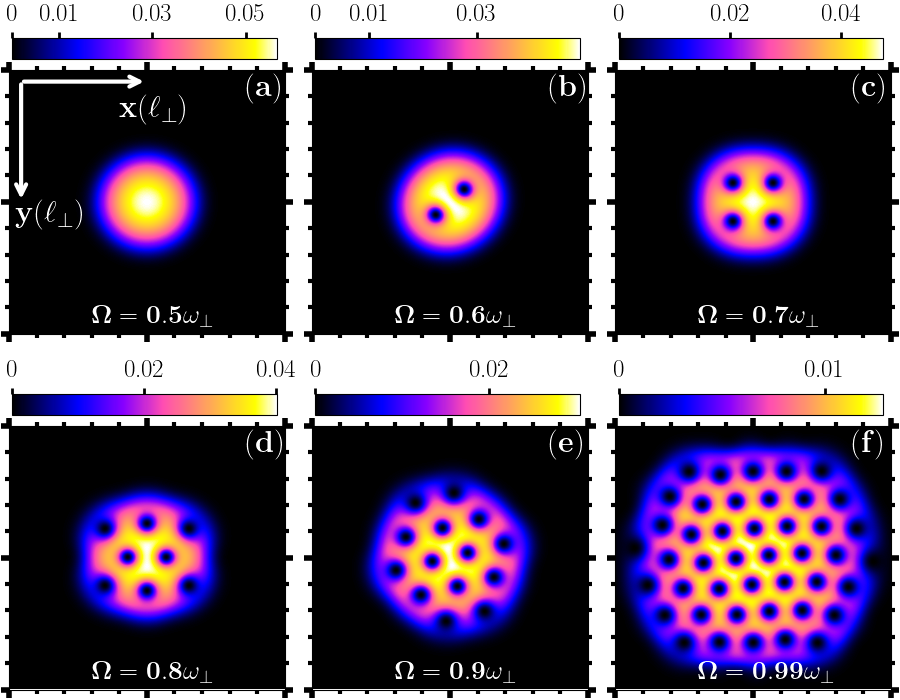}
\end{center}
\caption{(Color online) Density profiles $|\protect\psi |^{2}$ projected
onto the 2D plane, displaying vortex lattices with the nonlinearity
represented solely by the MF contact interactions with $g=100$. The rotation
frequences $\Omega $ (in units of $\protect\omega _{\perp }$) are indicated
in the panels, with the density levels coded by the by color bars. The $x$
and $y$ coordinates (in units of $\ell _{\perp }$) cover the interval $ 
[-10,+10]$. 
}
\label{fig02}
\end{figure}
The numerical results are exemplified by the vortex formation in two cases
when the MF nonlinear terms in the GP equation represent the contact and
dipolar interactions. The obtained results are consistent with previously
reported ones, such as those for the rotating systems with the pure contact
nonlinearity \cite{1999Butts}, and for the 3D dipolar rotating system with a
high value of the aspect ratio \cite{2016Kumar}. Here, it is not only illustrative
to consider these results, which are also useful to estimate
effects produced by the MF terms in comparison to those produced by the LHY
nonlinearity. For this purpose, we keep the same DDI and contact-interaction
parameters throughout the present study. Figure~\ref{fig02} shows how the
vortices emerge following the variation of the rotation frequency $\Omega $
in the usual nonlinear MF system, with $g=100$, in the absence of the DDI
and LHY terms ($g_{dd}=\eta =0$) in Eq.~\eqref{equ:gpe2d}.
Consistent with the critical frequency \eqref{critic} for the single vortex
in the 2D BEC trapped in the harmonic-oscillator potential with frequency $ 
\omega _{\perp }$ (where $\mu \approx 2.58$ is the scaled chemical
potential), it is observed that the generation of vortices starts slightly
below $\Omega \approx 0.6$, with the number of vortices $N_{v}$ increasing
almost linearly in the interval of $0.7<\Omega <0.9$. In Fig~\ref{fig03}, 
our focus is on the generation of vortices with the nonlinearity represented solely
by the DDI term, so that $(g,g_{dd},\eta )=(0,100,0)$. Comparing
the results displayed in Figs.~\ref{fig02} and ~\ref{fig03}, one can
understand the efficiency of the respective parameters for the control of
the production of vortices. These two cases, presented in Figs.~\ref{fig02}
and \ref{fig03}, in the absence of the LHY nonlinearity, will be useful for
comparison with the impact of the LHY term on the vortex production, which
is the subject of the next subsection.
\begin{figure}[th]
\begin{center}
\includegraphics[width=8.5cm]{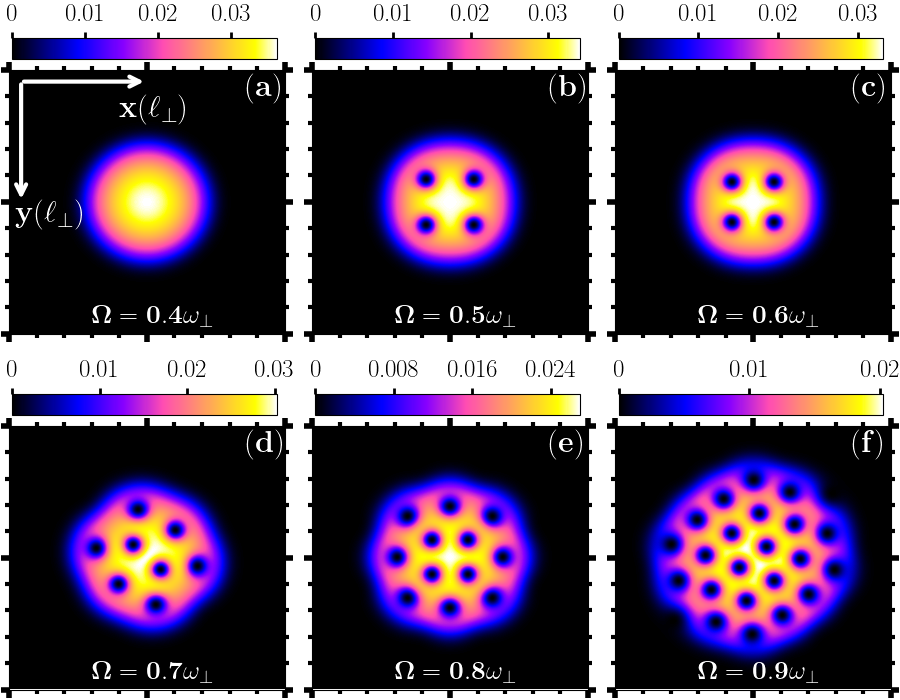}
\end{center}
\caption{(Color online) Density profiles $|\protect\psi |^{2}$ projected
onto the 2D plane, displaying vortex lattices with the nonlinearity
represented solely by DDI with $g_{dd}=100$. The rotation
frequencies $\Omega $ are indicated in the panels, with the density levels
coded by the color bars. The spatial domain and units are the same as in
Fig.~\protect\ref{fig02}. }
\label{fig03}
\end{figure}
\subsection{Vorticity in dipolar BEC with contact and LHY nonlinearities}
\label{secIIIB} The effect of LHY correction to the MF GP equation is
estimated in this section by assuming $\eta =200$ in Eq.\eqref{equ:gpe2d},
with $g=100$ and without dipolar interactions. By strictly considering the
expression \eqref{eta}, we should be aware that any specific value assumed
for $\eta $ will depend on $a_{s}$ and $a_{\mathrm{dd}}$. In particular,
when $a_{s}=0$, the quantum fluctuations become zero, such that $\eta\neq 0$
with $a_{s}=0$ has no interest in our present analyses. 
However, we can verify rotational effects due to quantum fluctuation in a non-dipolar BEC
system. For such a case, by assuming $g_{dd}=0$ in Eq.~ 
\eqref{equ:gpe2d}, we first consider the results obtained in Fig.~\ref{fig02}
with $g=100$ to estimate how strong the parameter $\eta $ of the quantum
fluctuations could be assumed to reach some visible effect in the results.
By looking for that, we have fixed $\eta =200$ in our following studies,
with the results being presented in Fig.~\ref{fig04}. The quantum
fluctuation contribution to the results, as expected, 
is quite small when compared with the MF nonlinear one, such that it starts 
to be noticeable only
with $\eta =200$ at $\Omega $ between 0.5 and 0.6. This can be verified in
our results of Figs.~\ref{fig02} and ~\ref{fig04}. The effect of quantum
fluctuations with such $\eta $ can also be noticed for large rotations,
close to $\Omega =0.99$.

\begin{figure}[th]
\begin{center}
\includegraphics[width=8.5cm]{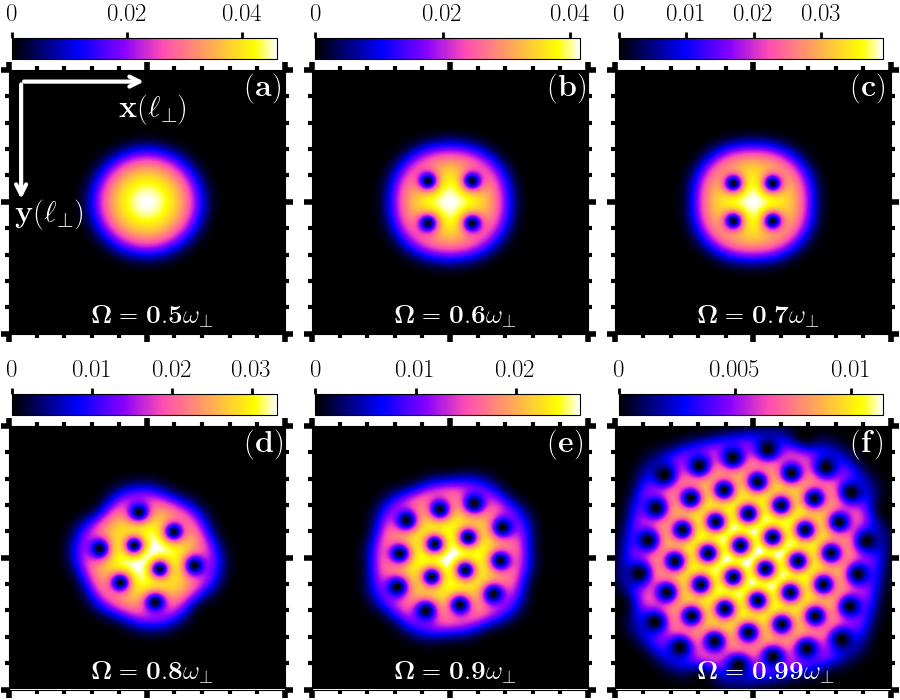}
\end{center}
\caption{(Color online) Density profiles $|\protect\psi |^{2}$ projected
onto the 2D plane, displaying vortex lattices with nonlinearity provided by
contact and LHY interactions, with $g=100$ and $\protect\eta=200$. The
frequencies are indicated inside the panels, with the respective density
levels given by color bars. The spatial domain and units are the same as in
Fig.~\protect\ref{fig02}. }
\label{fig04}
\end{figure}

With the given results in Fig.~\ref{fig05}, we are further verifying the
significance of the LHY correction with $\eta =200$. In this case, by
considering both contact and dipolar nonlinearities together with the
quantum fluctuations, one should notice that LHY quantum fluctuation with $ 
\eta =200$ remains producing a small effect. The main change in the observed
number of vortices in Fig.~\ref{fig05} is due to the cumulative effects
related to contact and dipolar interactions, as compared with previous
results shown in Figs.~\ref{fig02} and \ref{fig03}. Therefore, a possible
way to enhance the effect of LHY correction, in comparison with other
nonlinear effects, relies on playing with the signals of dipole-dipole and
contact interactions, which can be performed by noticing that the strength
of the LHY term, in the case of dipolar BECs, is only affected by the square
of $\varepsilon _{\mathrm{dd}}$, as shown by Eq.~\eqref{epsphi}. Therefore,
one can maximize or minimize the DDI effect just by controlling the $\varphi
$ dependence of $\eta $. This is being considered in the next section~\ref 
{secIIIC} as a possible mechanism to generate a pure-LHY dipolar superfluid.

\begin{figure}[th]
\begin{center}
\includegraphics[width=8.5cm]{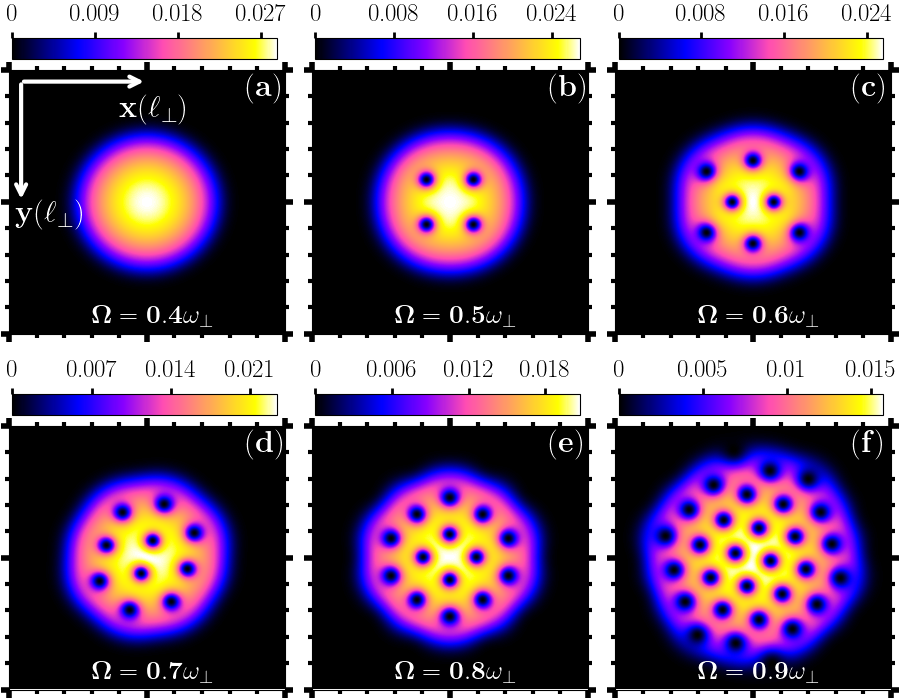}
\end{center}
\caption{(Color online) Density profiles $|\protect\psi |^{2}$ projected
onto the 2D plane, displaying vortex lattices with nonlinearity provided by
contact, dipolar and LHY interactions, with $g=100$, $g_{dd}=100$, and $ 
\protect\eta=200$. The frequencies are indicated inside the panels, with the
respective density levels given by color bars. The spatial domain and units
are the same as in Fig.~\protect\ref{fig02}.}
\label{fig05}
\end{figure}

\subsection{The generation of vortices in the pure-LHY superfluid}

\label{secIIIC} The scenario where the LHY correction is the 
only source of the nonlinearity in the single-component dipolar BEC is explored in this
subsection. It should be pointed out at this juncture that nonlinearity solely 
due to LHY was previously considered in the two-component BEC mixture with
contact interactions \cite{Jorgensen2018,Skov2021}. When the intra-species
repulsion and inter-species attraction are precisely balanced, the MF
contributions can be minimized or exactly canceled. To provide the
approximate cancellation of the nonlinear MF local and nonlocal terms in the
framework of the GP equation, we here consider the quasi-2D single-component
dipolar BEC system introduced in the previous section. In this scenario, one
can design the DDI intensity, by suitably adjusting the angle $\varphi$ of the
dipoles with respect to axis $z$ (as represented in Fig.~\ref{fig01}), and 
the contact interaction, through the scattering length $a_{s}$, by using the 
Feshbach resonance mechanism. Following that, within some approximation, we are 
left solely with the LHY nonlinearity, as postulated by Eq.~\eqref{EnLHY}, where 
the quintic term 
represents the LHY nonlinearity, with no explitic contribution from the contact 
and dipolar interactions.

\begin{figure}[th]
\begin{center}
\includegraphics[width=8.5cm]{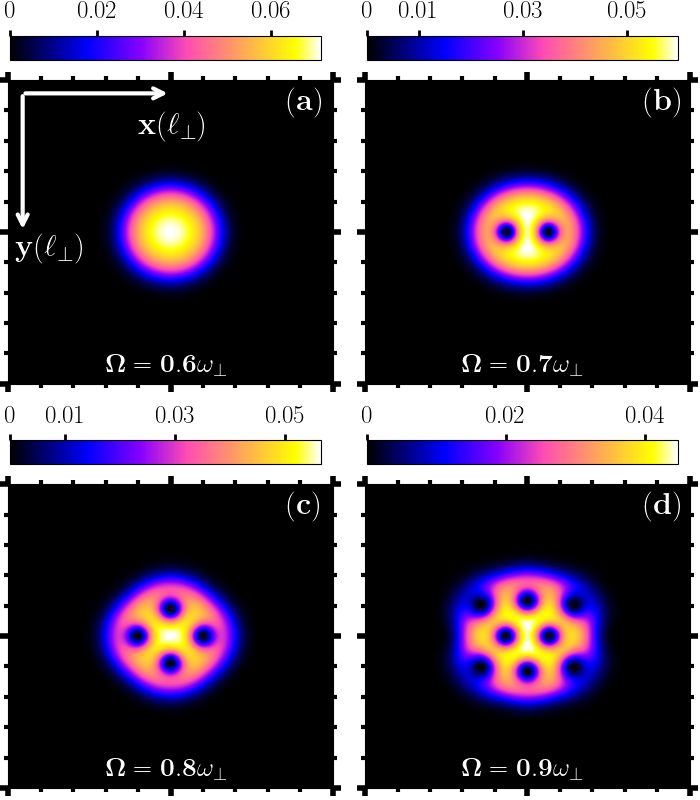}
\end{center}
\caption{(Color online) Density profiles $|\protect\psi |^{2}$ projected
onto the 2D plane, which display vortex lattices supported solely by the LHY
nonlinearity, with $\protect\eta =200$ in Eq. (\protect\ref{equ:gpe2d}). The
rotation frequencies $\Omega $ are indicated in the panels, with the
respective density levels defined by the color bars. The spatial domain and
units are the same as in Fig.~\protect\ref{fig02}. }
\label{fig06}
\end{figure}

The rotating dipolar LHY superfluid is exemplified by setting $\eta =200$ in
Eq.~\eqref{EnLHY}. For this dipolar BEC system, the values of $a_{s}$ and $ 
a_{\mathrm{dd}}$, which are included in expression (\ref{eta}), are not
critical, provided that they are tuned to secure the approximate
cancellation of the local and nonlocal MF terms. By changing the rotation
frequency $\Omega $ from $0$, by increments of $\Delta \Omega =0.1$, we have
observed no vortex produced for $\Omega \leq 0.6$ (in units of $\omega
_{\perp }$), with a pair of vortices being verified at $\Omega =0.7$. Next, two
and four pairs ($N_{v}=4$ and $8$) are observed at $\Omega =0.8$ and $\Omega
=0.9$, respectively. These results are shown in Fig.~\ref{fig06}. In more
detail, we explored the values of $\Omega $ to produce the first four
vortices, with results for the density plots and corresponding phases
presented in Fig.~\ref{fig07}. 
The phase map uses a color scale from $-\pi$ to $+\pi$, determined by the 
$\arctan({\rm Im}\psi /{\rm Re}\psi)$.
Beyond the trap boundaries, the phase is a numerical artifact, which 
arises from densities that are numerically small but not vanish.

As observed by the $\Omega$ values indicated in the four panels of 
Fig.~\ref{fig07}, particularly in the cases
to produce one and three stable vortices, careful fine tuning of 
$\Omega$ was required, with the values of $\Omega$ being, respectively, $0.6401$ and 
$0.768$. These findings reveal criticality in the states concerning
$\Omega$, for relatively small number of produced vortices, $N_{v}$. 
More explicitly, at $\Omega =0.6400$, no vortex can be found, whereas 
two vortices are generated at $0.6410\leq \Omega \leq 0.767$, 
as shown in Fig.~\ref{fig07}(b). 
Next, by slightly increasing the rotation frequency to 
$\Omega=0.768$, 
a state with three-vortex is obtained, which persists in a very 
narrow interval, $0.768\leq 0.778$. A small
variation in $\Omega $ can transform the three-vortex state into a two- or
four-vortex one (at $\Omega \leq 0.767$ and $\Omega \geq 0.779$,
respectively. The four-vortex state is produced in a broader interval, 
$0.779\leq \Omega \leq 0.881$, with Fig.~\ref{fig08}(d) displaying the 
result for $\Omega =0.779$.
This behavior, with vortex configurations being more stable with 
$2$ and $4$ vortices than with $1$ and $3$ vortices, is apparently a
characteristics of the pure LHY superfluid in our quasi-2D approach, 
which is represented by the quartic nonlinearity in the GP equation
with a strong pancake-like aspect ratio.

\begin{figure}[th]
\centering\includegraphics[width=0.95\linewidth]{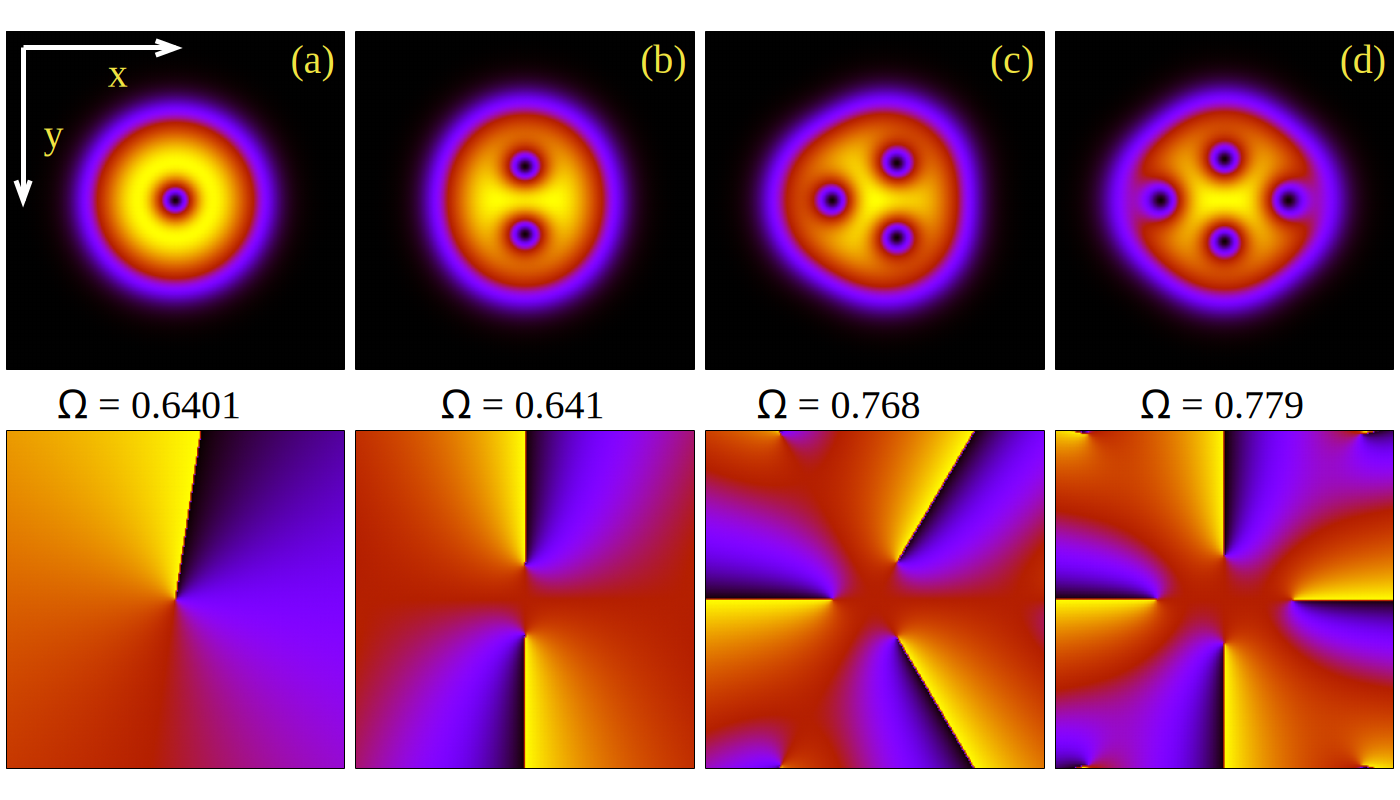} \centering 
\includegraphics[width=0.99\linewidth]{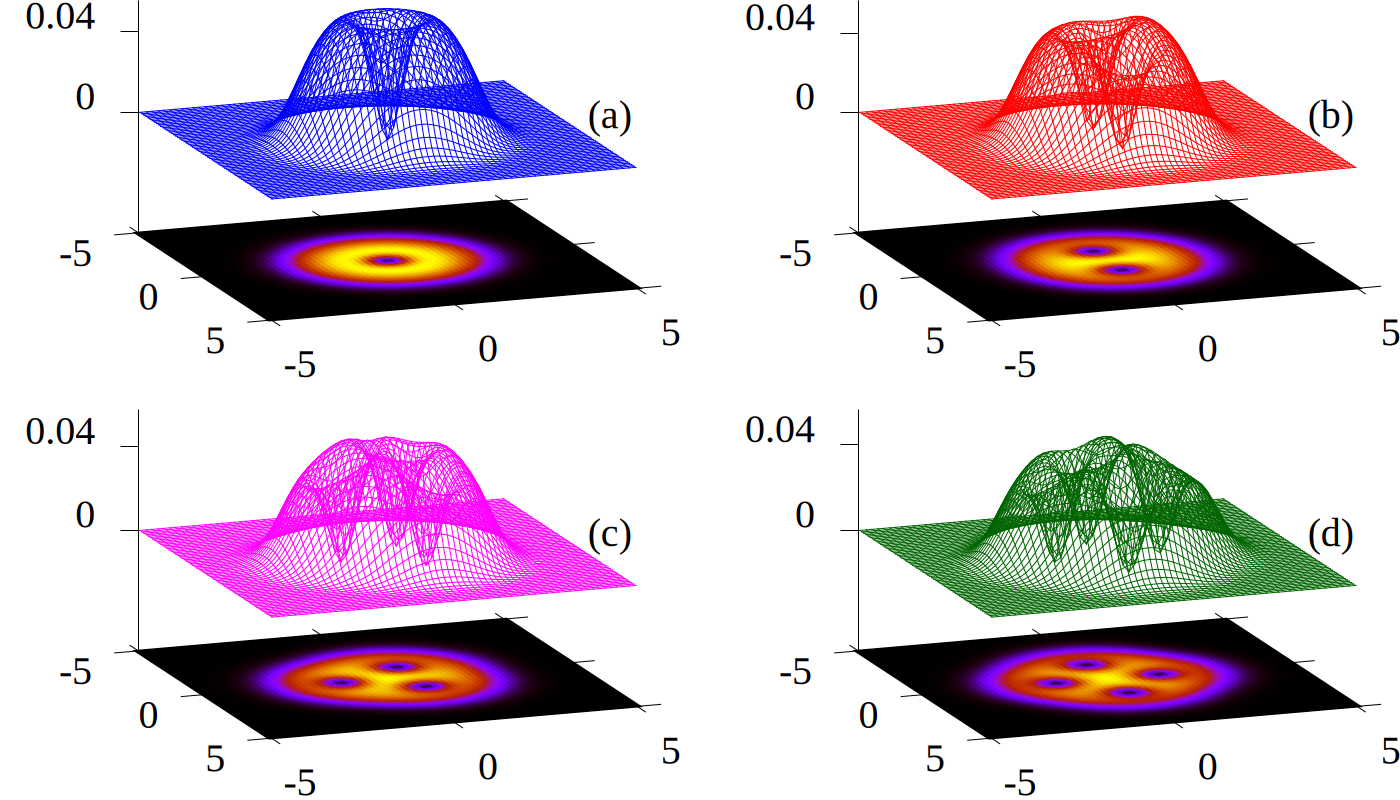}
\caption{(Color online) For the LHY superfluid with $\protect\eta =200$ at
four values of $\Omega $ (in units $\protect\omega _{\perp }$), the density
profiles $|\protect\psi |^{2}$ shown in the first row with the respective 
phases patterns shown in second
row  while the corresponding 3D representations drawn in the lower rows
 of the figure. Panel (a) pertains to the critical frequency, 
 $\Omega_{c}=0.6401$, for the creation of the single-vortex configuration.
The phase color map (the second row) spans from $-\pi$ (darkest blue) to $+\pi$ 
(brightest yellow), encircling each singularity in a clockwise direction, 
with zero value denoted by the maximum-red color.
}
\label{fig07}
\end{figure}

\begin{figure}[th]
\centering\includegraphics[width=0.95\linewidth]{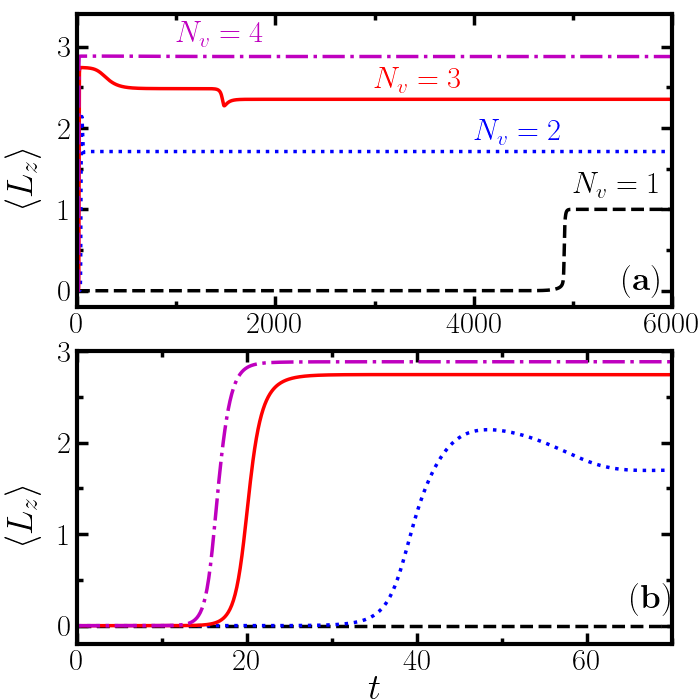}
\caption{(Color online) In the case when the nonlinearity is represented
solely by the LHY term, we show the long-time evolution of the angular
momentum $\langle L_{z}\rangle $ (in units of $\hbar $), corresponding to
the production of one to four vortices, as indicated in panel (a). In panel
(b) we display the same case as in (a) for a much shorter time interval, $ 
t<70$ (in units $\protect\omega _{\perp }^{-1}$). The respective values for
the rotation velocity $\Omega $ are given in the caption of
Fig.~\protect\ref{fig07}. The four established values, $\langle L_{z}\rangle 
= 1, 1.75, 2.35$, and $2.85$, agree with the ones reported in 
Ref.~\protect\cite{1999Butts} for the case of pure contact interactions.}
\label{fig08}
\end{figure}

Our results for the pure-LHY nonlinearity are consistent with results
reported in Ref.~\cite{1999Butts}, in which only the contact interactions
have been considered, and the predictions were produced for rotating states
within a more general 3D framework. For the first four stable vortices shown
in Figs.~\ref{fig07}, with $\Omega =0.6401,0.641,0.768$, and $0.779$, the
respective values of the system-averaged angular momentum per particle is $ 
\langle L_{z}\rangle =1,1.75,2.35$ and $2.85$. These results are analyzed in
Fig.~\ref{fig08}, considering the respective time evolution of $\langle
L_{z}\rangle $. Figure~\ref{fig08} shows how far one needs to evolve $\langle
L_{z}\rangle $ to obtain stable vortex states. In addition, as the
single-vortex configuration shown in (a$_{1}$) of Fig.~\ref{fig07} can only
be obtained after fine-tuning the rotational frequency within four decimal
digits, we can identify the critical rotation frequency for the
production of a single vortex in the pure-LHY case as $\Omega _{c}=0.6401$.
The corresponding numerical value of the chemical potential is 
$\mu_{\mathrm{QS}}=1.78$. This result is close enough to the one that can be
extracted from the TF relation~\eqref{critic}, $\mu \approx 1.86$, for the
same $\Omega_{c}$.

Usually, the vortex patterns are made energetically preferable states with
the help of the trap symmetry and boundary conditions. As shown in 
Refs.~\cite{1996Dalfovo,1996Baym,1999Butts}, for contact interactions, the critical
rotation frequencies $\Omega _{cn}$, which are necessary to attain stable
states with $n$ vortices are determined by the MF two-body interaction.
Therefore, for vortices supported by the pure LHY nonlinearity, such
critical frequencies and the corresponding stability may be estimated by
means of the respective coefficient $\eta$, given in Eq. (\ref{eta}).

Figure~\ref{fig09} summarizes the results for $N_{v}$ versus $\Omega $, 
varying $\Omega $ up to $0.99$, for the solely-LHY nonlinearity, 
with strength $\eta =200$. In this case, the largest number of
vortices found for $\Omega =0.99$ is $N_{v}=20$.
The figure displays the range intervals of 
frequencies $\Delta\Omega$ for the given specific values of $N_v$.  
Notable are the cases with $N_{v}=2, 4$ and $8$, which have larger
$\Delta\Omega$ than the other cases. Also, noticeable is the case with
$N_v=1$, which has $\Omega_c=0.6401$ with $\Delta\Omega<0.001$
(In this regard, see also the upper row of Fig.\ref{fig07}).
Inside Fig.~\ref{fig09}, we have also included density plots obtained 
for $N_{v}=5,6,7$ and $10$, to complement  the ones previously  
shown for $N_{v}\leq 4$.

\begin{figure}[th]
\centering\includegraphics[width=0.99\linewidth]{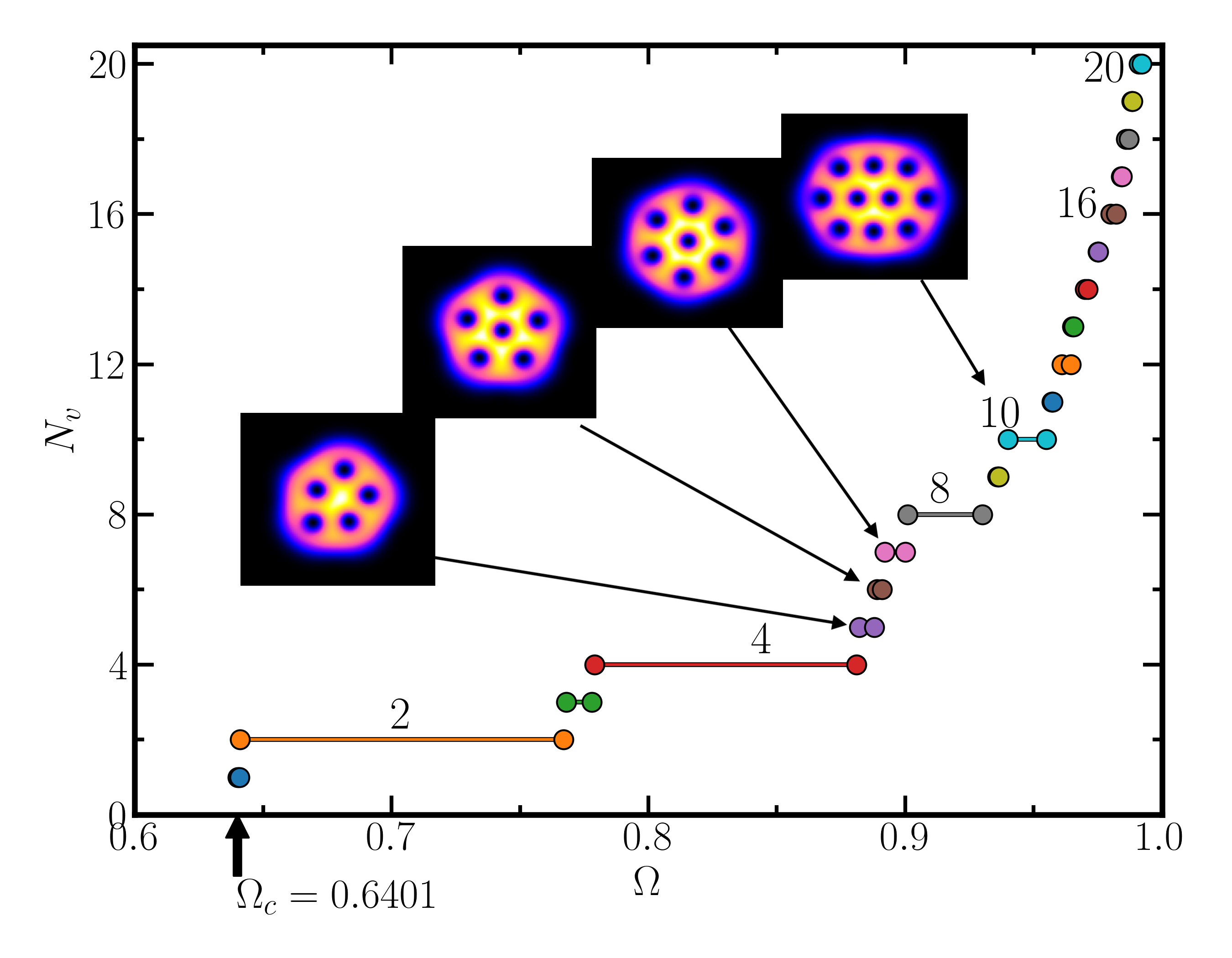}
\caption{The dependence of the number of vortices, $N_{v}$, on the rotation
frequency $\Omega $ (in units of $\protect\omega_\perp $) for the pure LHY
case, with $\protect\eta =200$ (see Eq. (\protect\ref{eta})). The position
of the single-vortex critical frequency, $\Omega _{c}=0.6401$, is indicated.
For even numbers $N_{v}\geq 2$, the existence ranges of $\Omega $ are
indicated too. The density profiles for $N_{v}=5,6,7,$ and $10$ are plotted
in the panel. }
\label{fig09}
\end{figure}

\subsection{The effect of the rotational frequency}

For the usual BEC system trapped in the harmonic oscillator potential, with
the cubic MF nonlinearity, the rotating states were first predicted in 
Ref.~\cite{1999Butts} in the framework of a more general 3D framework,
considering the generation of vortices, with the corresponding angular
momentum being a function of the rotation velocity $\Omega $. The previous
section, confirms the occurrence of the same kind of vortex-generating
mechanism in the LHY-dominated condensate, as in the case of the one with the
MF nonlinearity. By adding the rotation to the system, all vortices are
produced with the same vorticity sign, without vortex-antivortex
annihilation.

{\footnotesize
\begin{table*}[t]
\caption{The effect of rotation frequency $\Omega $ on chemical potential ($ 
\protect\mu $), total energy per particle ($E$), number of visible vortices 
($N_{v}$), and angular momentum per particle, $\left\langle
L_{z}\right\rangle $ for five different sets of parameters considered in
this work. Case I (the pure MF contact interaction, with $g=100$); case II
(the pure DDI, with $g_{dd}=100$); case III (with the contact-MF
and LHY terms, without DDI, for $g=100$ and $\protect\eta =200$); case IV
(the nonzero contact-MF, DDI, and LHY terms, with $g=100$, 
$g_{dd}=100$, and $\protect\eta=200$); and case V 
(the pure LHY nonlinearity, with $\protect\eta=200)$. 
In the pure-LHY superfluid, one should observe that the critical frequencies 
$\Omega_c$ for producing one and two vortex states are about the same within two
decimal digits (being, respectively, $0.6401$ and $0.6410$).
It implies that a single-vortex production is numerically
quite challenging. In this table, the corresponding values
for $N_v=1$ are: $\Omega=0.6401$, $\mu=1.78$, $E=1.21$, and $\langle L_z\rangle=1$.
}
\label{tab1}
\begin{center}
{\small
\begin{tabular}{|c|ccccc|ccccc|ccccc|ccccc|}
\hline\hline
\multirow{2}{*}{$\Omega$} 
& \multicolumn{5}{c|}{$\mu$ ($\hbar\omega_\perp$)}
& \multicolumn{5}{c|}{$E$ ($\hbar\omega_\perp$)} 
& \multicolumn{5}{c|}{$N_v$}
& \multicolumn{5}{c|}{$\langle L_z\rangle$ ($\hbar$)} \\
\cline{2-21}\cline{2-21}
& I & II & III & IV & V & I & II & III & IV & V & I & II & III & IV & V & I
& II & III & IV & V \\ \hline\hline
0.0 & 2.96 & 4.61 & 3.40 & 5.62 & 2.06 & 2.05 & 3.10 & 2.30 & 3.76 & 1.40 & 0
& 0 & 0 & 0 & 0 & 0 & 0 & 0 & 0 & 0 \\
0.4 & 2.96 & 4.61 & 3.40 & 5.62 & 2.06 & 2.05 & 3.10 & 2.30 & 3.76 & 1.40 & 0
& 0 & 0 & 0 & 0 & 0 & 0 & 0 & 0 & 0 \\
0.5 & 2.96 & 4.12 & 3.40 & 4.99 & 2.06 & 2.05 & 2.68 & 2.30 & 3.18 & 1.40 & 0
& 4 & 0 & 4 & 0 & 0 & 2.76 & 0 & 2.99 & 0 \\
0.6 & 2.59 & 3.80 & 2.83 & 4.37 & 2.06 & 1.76 & 2.42 & 1.84 & 2.77 & 1.40 & 2
& 4 & 4 & 8 & 0 & 1.69 & 3.06 & 2.72 & 4.68 & 0 \\
0.7 & 2.24 & 3.07 & 2.50 & 3.54 & 1.61 & 1.48 & 1.89 & 1.58 & 2.16 & 1.09 & 4
& 8 & 4 & 10 & 2 & 2.85 & 4.91 & 3.05 & 6.13 & 1.71 \\
0.8 & 1.62 & 2.22 & 1.69 & 2.51 & 1.23 & 0.99 & 1.23 & 0.96 & 1.26 & 0.80 & 8
& 12 & 8 & 14 & 4 & 4.34 & 6.98 & 5.00 & 8.44 & 2.88 \\
0.9 & 0.38 & 0.24 & 0.47 & 0.04 & 0.55 & -0.05 & -0.47 & -0.07 & -0.03 & 0.24
& 14 & 22 & 14 & 28 & 8 & 7.19 & 11.38 & 8.10 & 14.12 & 4.53 \\
0.99 & -3.81 & -6.06 & -4.30 & -7.37 & -1.79 & -3.99 & -6.37 & -4.51 & -7.76
& -1.89 & 42 & 68 & 48 & 80 & 20 & 20.75 & 30.94 & 22.73 & 36.53 & 11.99 \\
\hline\hline
\end{tabular}
}
\end{center}
\end{table*}
}

The findings for the vortex configurations are summarized in Table~\ref{tab1}
for all five nonlinearity cases that we have considered. In the table, we report the
numerical results obtained for the chemical potential, total energy per
particle $E$, and number of visible vortices $N_{v}$, for specific values of
the rotation frequency $\Omega $, which are given with one decimal digit.
For the five cases being considered, the corresponding behaviors of $E$ and $ 
\mu$, as functions of $\Omega $, are presented in two panels of Fig.~\ref{fig10}. 
The energy cost of generating vortices can be verified from the
results provided in Table~\ref{tab1}. In particular, by focusing on case V
of the table, where we have only the LHY nonlinear term, the energy cost of
producing a pair of vortices, as compared with the non-rotating case 
($E_{\Omega =0}$), is $\approx 22\%$ at $\Omega =0.7$. On the other hand, when
we have only the MF contact nonlinearity (column I in the table), the cost
is $\approx 14\%$ at $\Omega =0.6$. This indicates that the vortices are
more likely to be produced in the presence of the pure MF contact
nonlinearity than in the case of the pure LHY nonlinearity.

\begin{figure}[th]
\centering
\includegraphics[width=0.99\linewidth]{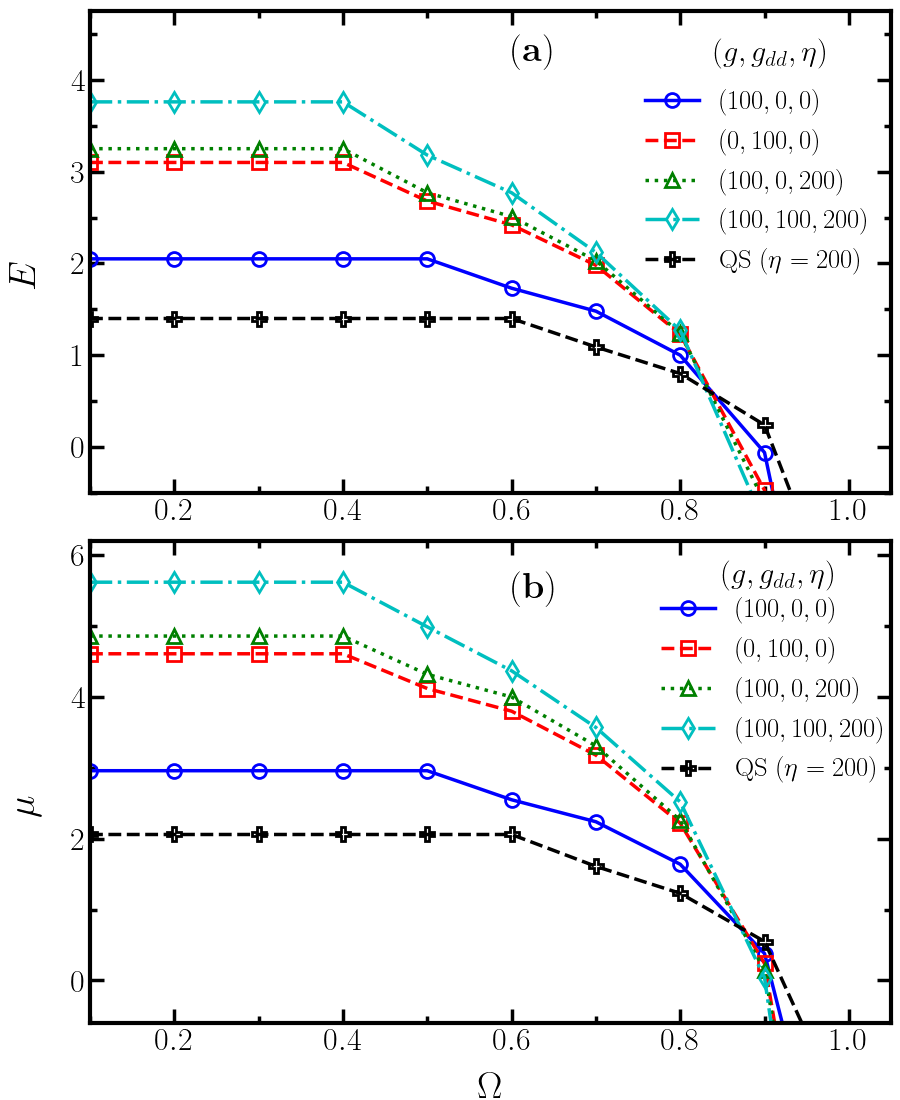}
\caption{(Color online) Energy $E$ per particle (a) and the corresponding
chemical potential $\protect\mu $ (b) are shown as functions of $\Omega $,
for five different combinations of the interactions considered in Table
\protect\ref{tab1} (as indicated in the panels). The energy and frequency
units are, respectively, $\hbar \protect\omega_{\perp }$ and $\protect 
\omega _{\perp }$. }
\label{fig10}
\end{figure}

\begin{figure}[th]
\centering\includegraphics[width=0.99\linewidth]{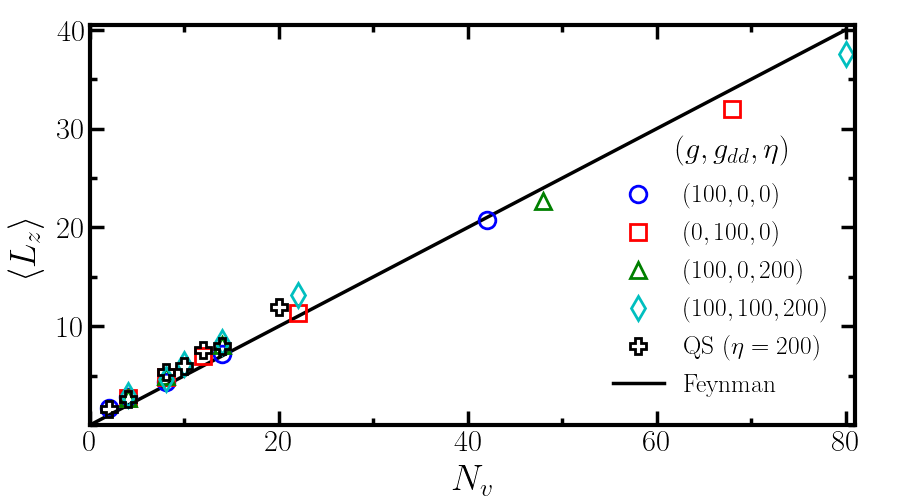}
\caption{(Color online) The linear increase of the angular momentum per
particle $\langle L_{z}\rangle $ (in units of $\hbar $), associated with the
change in the number of vortices, $N_{v}$, is confirmed for all the cases
presented in Table \protect\ref{tab1}. As shown, the results are close to
the linear behavior (solid line) predicted by Feynman~\protect\cite 
{1955Feynman}. The parameters are indicated in the panel, where QS stands
for the solutions corresponding to Eq. \eqref{EnLHY}, pertaining to the
pure-LHY nonlinearity.}
\label{fig11}
\end{figure}
Similarly, as shown in Ref.~\cite{2001Fetter} for the contact MF nonlinear
interaction, the critical frequency necessary to produce the first vortex is
given by the approximate analytical expression (\ref{critic}). Essentially,
such an approximate expression in terms of the chemical potential is
expected to be similar for all five cases considered in Table \ref{tab1}.
Given the exact numerical results for $\mu (\Omega )$, when starting to
deviate from the no-rotation case, $\mu (\Omega =0)$, we can obtain the
corresponding single-vortex critical frequencies. Roughly, these critical
frequencies can be numerically extracted from the results provided in 
Fig.~\ref{fig10}, as well as from Table~\ref{tab1}.

Figure~\ref{fig11} summarizes the relationship between the angular momentum per particle 
and the number of vortices $N_v$. The data exhibit a clear linear trend, in excellent 
agreement with the Feynman's prediction for the angular momentum in a rapidly rotating 
superfluid~\cite{1955Feynman}. The vortex core (where the density is zero) is a 
nodal line of the wave function. Preventing the kinetic energy from diverging, the 
core provides a cutoff, with a radius on the order of the healing length
$\xi=\hbar/\sqrt{2m n g_{eff}}$, where $g_{eff}$ is the effective interaction strength.
Next, we briefly discuss how the different kind of nonlinear interactions 
affect the corresponding healing lengths.

\subsection{The healing length in the LHY-corrected case}
From the density results presented in Figs.~\ref{fig02} and 
\ref{fig06}-\ref{fig07}, by comparing cases with identical numbers of vortices, 
one can verify that the vortex healing lengths (radial sizes) slightly increase
when proceeding from the usual cubic (pure contact) to the pure quartic LHY 
nonlinearity in the GP formalism.  
We can estimate the healing lengths more precisely using the results in 
Table~\ref{tab1}. As shown in Appendix A, the full-dimensional healing length 
for the LHY-corrected case is given by
$\xi={\hbar}/{\sqrt{2m\left(g_0 n_0 +\frac{3}{2}g_{LHY} n_0^{3/2}\right)}}$, where 
$n_0$ is the equilibrium bulk density. 
So, for the usual MF, we have 
$\xi_{MF}={\hbar}/{\sqrt{2m\left(g_0 n_0\right)}} = {\hbar}/{\sqrt{2m\mu_{MF}}}$; 
whereas, for pure LHY interaction,
$\xi_{LHY}={\hbar}/{\sqrt{3m\;g_{LHY} n_o^{3/2} }} = {\hbar}/{\sqrt{3m\mu_{LHY}}}$.
Using the dimensionless chemical potentials from Table~\ref{tab1}, we next compare 
the four-vortex configurations, where the vorticity is the most stable.
For the pure contact interaction: $\Omega=0.7$ (Fig.~\ref{fig02}c), with
$\mu_{MF}\approx 2.24$; and, for the pure LHY, $\Omega=0.8$ (Fig.~\ref{fig06}c), with
$\mu_{LHY}\approx 1.23$. Therefore, 
$\xi_{LHY}/\xi_{MF} \approx \sqrt{2\mu_{MF}/(3\mu_{LHY})}\approx 1.10$, which 
indicates that the vortices in pure LHY quantum fluids have healing lengths about 10\% 
larger than for the cases with only contact interactions. 
With the healing length slightly larger in the LHY fluid, the space between the radial
borders of the vortices will be reduced accordingly.
The intervortex distance (between the central cores), as shown in Ref.~\cite{2016Pitaevskii},
 is given by $l_\Omega=\sqrt{\hbar/(m\Omega)}$, which depends on the rotational frequency, 
 consistent with the Feynman's prediction~\cite{1955Feynman}. Therefore, 
when comparing cases having the same number of vortices but 
different nonlinearities, this intervortex distance should
be slightly changed according to the changes of the rotational frequency.
In our example of four vortices, the estimated distance between vortices is 
$l_{\Omega_{LHY}}/l_{\Omega_{MF}} \approx \sqrt{7/8}\approx 0.935$.

\section{Conclusion}

In the present work, considering the rotating dipolar BEC, we investigate
the impact of quantum fluctuations, represented by the LHY correction to the
GP equation, on the formation and dynamics of individual vortices and vortex
lattices. The main outcome of our analysis is related to the scenario of 
vortex production when the only source of nonlinearity is provided by the 
LHY term. This condition may approximately hold for the
single-component dipolar BEC, with the mean-field (MF) contact interaction
and DDI (dipole-dipole interaction) roughly canceling each other. To achieve
this setting, one has to select appropriate values of the strengths of the
MF contact and dipole-dipole interactions, which can be provided through the
Feshbach resonance, and by adjusting the angle that determines the
orientation of the atomic magnetic moments. It is shown that the 
LHY-only nonlinearity can be made strong enough to produce vortices in the dipolar
superfluid. For the smaller numbers of stable vortices ($N_{v}\leq 5$), it
being more likely to obtain even rather than odd numbers of vortices when
varying the rotational frequency $\Omega $. This means that the range of $ 
\Omega $ is relatively large to generate even numbers ($2$ or $4$) of the
vortices, being much smaller for the odd numbers ($1$ and $3$). This feature
is particularly pronounced in the pure LHY case, so that, for the
single-vortex case, the corresponding range of the rotation frequency $ 
\Omega $ is so narrow that it can be identified with the critical one $ 
\Omega_{c}$ for the production of the single vortex. This effect, related
to the quartic LHY\ nonlinearity, can also be partially attributed to the
aspect ratio of the 3D trap. In a less pronounced form, a similar effect was
also noted for the relatively small values of vortices in the case of pure
MF contact interactions. A more detailed investigation related to this
effect, considering different parametrization, may be required to clarify
the role of the aspect ratio and its interplay with the nonlinearity.

We have also computed, within the framework of our quasi-2D model, the
number of vortices, chemical potential, and energy per particle, expressed
as functions of the rotation frequency $\Omega $. For that purpose, we have
considered specific combinations of the strengths of the MF contact
interaction, DDI, and LHY nonlinearity. These parameters are assumed to be
large enough to study the vortex production from the perspective of possible
experimental realizations in dipolar BEC systems. The results demonstrate
that, while the vortex numbers consistently increase with the growth of $ 
\Omega $, the chemical potential and energy decrease. From this study, when
considering the pure LHY nonlinearity, 
exemplified by a particular value of the strength, 
which can be further generalized, we conclude how to precisely
identify the single-vortex critical frequency, along with the corresponding
values of the chemical potential and energy. In all the investigated cases, 
we have observed results consistent with 
Feynman's prediction of a linear dependence between  
angular momentum per particle and the number of vortices.
 Finally, we briefly discuss the vortex distribution and corresponding 
healing lengths in a trap in view of the different kinds of nonlinearities, by confronting 
the case with pure MF, with that one with pure LHY quantum fluid, with rotations 
such that the same number of vortices is produced. The results are 
consistent with experimental observations that $\xi_{LHY}$  $> \xi_{MF}$.

\begin{acknowledgments}
\noindent
S.S. thanks Funda\c c\~ao de Amparo \`a Pesquisa do Estado de S\~ao Paulo
(FAPESP) for the fellowship support, Proj.~2024/04174-8. R.R. wishes to acknowledge
the financial assistance from DST-CURIE(DST-CURIE/PG/54) and ANRF( CRG/2023/008153)
L.T. also acknowledges partial support from FAPESP (Proj. 2024/01533-7) and
Conselho Nacional de Desenvolvimento Cient\'\i fico e Tecnol\'ogico (CNPq)
(Proc. 303263/2025-3). 
\end{acknowledgments}

\setcounter{section}{0}
\setcounter{equation}{0}

\section*{Appendix: The healing length of a vortex in an LHY-corrected superfluid} 
\renewcommand\theequation{\thesection A\arabic{equation}}
The healing length for a vortex in an LHY-corrected superfluid is defined by the 
linearized density 
recovery around the bulk equilibrium, governed by an effective coupling $g_{eff}$, that
includes the derivative of the LHY term with respect to density. 
In view of our purpose, to simplify this appendix, we consider a non-dipolar system.
The density is zero at the core of the vortex and heals to the equilibrium density 
$n_0\equiv|\psi_0({\bf r})|^2$ as going away from it. This equilibrium density itself 
is determined by balancing the MF (with interaction $g_0$) and LHY (with interaction 
strength $g_{LHY}$) pressures in the bulk, 
often leading to a flat-top droplet if the confinement is weak. 
By ignoring the external potential, considering uniform gas bulk
equilibrium condition, the  pressure balance gives
\begin{equation}\label{ap01}
\mu_0 = g_0 n_0 + g_{LHY}{n_0}^{3/2}.
\end{equation}
For a stationary straight vortex line with winding number one, using the ansatz $\psi({\bf r})=\sqrt{n(r)}{\rm e}^{{\rm i}\varphi}$,
with $r$ being the distance from the core, the real part of the extended GP equation is 
{\small
\begin{equation}\label{ap02}
\hspace{-0.2cm}\frac{-\hbar^2}{2m}
\left[\frac{1}{r \sqrt{n}}\frac{d}{dr} \left(r\frac{d\sqrt{n}}{dr}\right)-\frac{1}{r^2}
\right] + g_0 n + g_{LHY}{n}^{3/2} = \mu_0.
\end{equation}
}At $r\to\infty$, $n\to n_0$, with small density deviations being $\delta n = n_0-n$, the
linearization yields the healing length.
The effective nonlinearity is now $g_0 n+g_{LHY} n^{3/2}$. By expanding the chemical
potential about $n_0$, we have
\begin{equation}\label{ap03}
\mu(n) \approx \mu_0+ \left( g_0 + \frac{3}{2}g_{LHY} n_0^{1/2}\right)(n-n_0),
\end{equation}
such that the effective coupling strength governing the stiffness against density variations is 
\begin{equation}\label{ap04}
g_{eff} =  g_0 + \frac{3}{2}g_{LHY} n_0^{1/2}.
\end{equation}
For a self-bound droplet, where $g_0$ can be negative, but $g_{eff}$ is positive 
and nonsmall, the LHY term provides the dominant repulsion stabilizing the core. 
In analogy with the standard formalism, 
the healing length of a vortex in an LHY-corrected superfluid is given by 
\begin{equation}\label{ap05}
\xi = \frac{\hbar}{\sqrt{2mg_{eff} n_0}} =  \frac{\hbar}{\sqrt{{2m } \left(g_0+\frac{3}{2}g_{LHY} n_0^{1/2}\right) n_0}} .
\end{equation} 
Therefore, in a regime dominated by the LHY term (the droplet regime), assuming $g_0=0$, we obtain
the healing length, as
\begin{equation}\label{ap06}
\xi_{LHY}\approx \frac{\hbar}{\sqrt{3m g_{LHY} n_0^{3/2}}} \;\;\; = \;\;\; \frac{\hbar}{\sqrt{3m \mu_{LHY}}},
\end{equation}
which scales with the bulk density as $n_0^{-3/4}$, different from the MF scaling $\sim n_0^{-1/2}$.
In terms of the respective chemical potentials, the factor
$1/\sqrt{2}$ is replaced by $1/\sqrt{3}$.}


\begin{thebibliography}{999}
\bibitem{Griesmaier:2006} A. Griesmaier, J. Stuhler, T. Koch, M. Fattori, T.
Pfau, and S. Giovanazzi, Comparing contact and dipolar interactions in a
Bose-Einstein condensate, Phys. Rev. Lett. \textbf{97}, 250402 (2006).

\bibitem{Lahaye:2007} T. Lahaye, T. Koch, B. Fr\"ohlich, M. Fattori, J.
Metz, A. Griesmaier, S. Giovanazzi and T. Pfau, Strong dipolar effects in a
quantum ferrofluid, Nature \textbf{448}, 672 (2007).

\bibitem{Koch:2008} T. Koch, T. Lahaye, J. Metz, B. Fr\"ohlich, A.
Griesmaier and T. Pfau, Stabilization of a purely dipolar quantum gas
against collapse, Nature Phys. \textbf{4}, 218 (2008).

\bibitem{Youn:2010} S. H. Youn, M. Lu, U. Ray, and B. L. Lev, Dysprosium
magneto-optical traps, Phys. Rev. A \textbf{82}, 043425 (2010).

\bibitem{Lu:2011} M. Lu, N. Q. Burdick, S. H. Youn, and B. L. Lev, Strongly
dipolar Bose-Einstein condensate of dysprosium, Phys. Rev. Lett. \textbf{107}
, 190401 (2011).

\bibitem{Aikawa:2012} K. Aikawa, A. Frisch, M. Mark, S. Baier, A. Rietzler,
R. Grimm, and F. Ferlaino, Bose-Einstein condensation of erbium, Phys. Rev.
Lett. \textbf{108}, 210401 (2012).

\bibitem{dbec1} T. Lahaye, C. Menotti, L. Santos, M. Lewenstein, T. Pfau,
The physics of dipolar bosonic quantum gases, Rep. Prog. Phys. \textbf{72},
126401 (2009).

\bibitem{dbec2} M. A. Baranov, M. Dalmonte, G. Pupillo, and P. Zoller,
Condensed matter theory of dipolar quantum gases Chem. Rev. \textbf{112},
5012 (2012).

\bibitem{2002Giovanazzi} S. Giovanazzi, A. G\"orlitz, and T. Pfau, Tuning
the dipolar interaction in quantum gases, Phys. Rev. Lett. \textbf{89},
130401 (2002).

\bibitem{Feshbach2010} C. Chin, R. Grimm, P. Julienne, and E. Tiesinga,
Feshbach resonances in ultracold gases, Rev. Mod. Phys. \textbf{82}, 1225
(2010).

\bibitem{2019Kumar} R. K. Kumar, L. Tomio, and A. Gammal, Spatial separation
of rotating binary Bose-Einstein condensates by tuning the dipolar
interactions, Phys. Rev. A \textbf{99}, 043606 (2019).

\bibitem{2003ODell} D. H. J. O'Dell, S. Giovanazzi and G. Kurizki, Rotons in
gaseous Bose-Einstein condensates irradiated by a laser, Phys. Rev. Lett.
\textbf{90}, 110402 (2003).

\bibitem{1955Feynman} R. P. Feynman, in Progress in Low Temperature Physics
vol 1, ed. C. J. Gorter (Amsterdam: North-Holland, 1955) p 17.

\bibitem{Noziers} P. Nozi\`{e}rez, Is the roton in superfluid 4He the ghost
of a Bragg spot?, J. Low Temp. Phys. \textbf{137}, 45-67 (2004).

\bibitem{Santos} L. Santos, G. V. Shlyapnikov, and M. Lewenstein,
Roton-Maxon Spectrum and Stability of Trapped Dipolar Bose-Einstein
Condensates, Phys, Rev. Lett. \textbf{90}, 250403 (2003).

\bibitem{Fisher} U. R. Fischer, Stability of quasi-two-dimensional
Bose-Einstein condensates with dominant dipole-dipole interactions, Phys.
Rev. A \textbf{73}, 031602(R) (2006).

\bibitem{Bohn} C. Ticknor, R. M. Wilson, and J. L. Bohn, Anisotropic
Superfluidity in a Dipolar Bose Gas, Phys. Rev. Lett. \textbf{106}, 065301
(2011).

\bibitem{rotons-new} A. Villois, M. Onorato, and D. Proment, Vortex to
Rotons Transition in Dipolar Bose-Einstein Condensates, Phys. Rev. Lett.
\textbf{124}, 253401 (2025).

\bibitem{Berge1998} L. Berg\'{e}, {Wave collapse in physics: principles and
applications to light and plasma waves}, Phys. Rep. \textbf{303}, 259 (1998).

\bibitem{Sulem} C. Sulem and P.-L. Sulem, \textit{The nonlinear Schr\"{o} 
dinger equation: Self-Focusing and wave collapse} (Springer, New York, 1999).

\bibitem{2001Gammal} A. Gammal, T. Frederico, and L. Tomio, Critical number
of atoms for attractive Bose-Einstein condensates with cylindrically
symmetrical traps, Phys. Rev. A \textbf{64}, 055602 (2001).

\bibitem{Fibich} G. Fibich, The nonlinear Schr\"{o}dinger equation: Singular
solutions and optical collapse (Springer, Heidelberg, 2015).

\bibitem{2020Kartashov} Y. V. Kartashov, L. Torner, M. Modugno, E. Ya.
Sherman, B. A. Malomed and V. V. Konotop, Multidimensional hybrid
Bose-Einstein condensates stabilized by lower-dimensional spin-orbit
coupling Phys. Rev. Res. \textbf{2}, 013036 (2020).

\bibitem{2022-Malomed-Book} B. A. Malomed, \textit{Multidimensional Solitons}
(AIP Publishing, Melville, New York, 2022).

\bibitem{Altmeyer2007} A. Altmeyer, S. Riedl, C. Kohstall, M. J. Wright, R.
Geursen, M. Bartenstein, C. Chin, J. H. Denschlag, and R. Grimm, Precision
measurements of collective oscillations in the BEC-BCS crossover, Phys. Rev.
Lett. \textbf{98}, 040401 (2007).

\bibitem{Shin2008} Y.-i. Shin, A. Schirotzek, C. H. Schunck, and W.
Ketterle, Realization of a strongly interacting Bose-Fermi mixture from a
two-component Fermi gas, Phys. Rev. Lett. \textbf{101}, 070404 (2008).

\bibitem{Papp2008} S. B. Papp, J. M. Pino, R. J. Wild, S. Ronen, C. E.
Wieman, D. S. Jin, and E. A. Cornell, Bragg spectroscopy of a strongly
interacting $^{85}$Rb Bose-Einstein condensate, Phys. Rev. Lett. \textbf{101} 
, 135301 (2008).

\bibitem{Petrov2015} D. S. Petrov, Quantum mechanical stabilization of a
collapsing Bose-Bose mixture, Phys. Rev. Lett. \textbf{115}, 155302 (2015).

\bibitem{Petrov2016} D. S. Petrov and G. E. Astrakharchik, Ultradilute
low-dimensional liquids, Phys. Rev. Lett. \textbf{117}, 100401 (2016).

\bibitem{Lee1957} T. D. Lee and C. Yang, Many-Body problem in quantum
mechanics and quantum statistical mechanics, Phys. Rev. \textbf{105}, 1119
(1957).

\bibitem{Lee1957a} T. D. Lee, K. Huang, and C. N. Yang, Eigenvalues and
eigenfunctions of a Bose system of hard Spheres and its low-temperature
properties, Phys. Rev. \textbf{106}, 1135 (1957).

\bibitem{2002Bulgac} A. Bulgac, Dilute quantum droplets, Phys. Rev. Lett.
\textbf{89}, 050402 (2002).

\bibitem{1973Efimov} V. Efimov, Energy levels of three resonantly
interacting particles, Nucl. Phys. A \textbf{210}, 157 (1973).

\bibitem{2000Gammal} A. Gammal, T. Frederico, L. Tomio and Ph. Chomaz,
Atomic Bose-Einstein condensation with three-body interactions and
collective excitations, J. Phys. B: At. Mol. Opt. Phys. \textbf{33}, 4053
(2000).

\bibitem{Feshbach} G. Roati, M. Zaccanti, C. D'Errico, J. Catani, M.
Modugno, A. Simoni, M. Inguscio, and G. Modugno, 39K Bose-Einstein
Condensate with Tunable Interactions, Phys. Rev. Lett. \textbf{99}, 010403
(2007).

\bibitem{Jorgensen2018} N. B. J\o rgensen, G. M. Bruun, and J. J. Arlt,
Dilute fluid governed by quantum fluctuations, Phys. Rev. Lett. \textbf{121} 
, 173403 (2018).

\bibitem{Skov2021} T. G. Skov, M. G. Skou, N. B. J\o rgensen, and J. J.
Arlt, Observation of a Lee-Huang-Yang fluid, Phys. Rev. Lett. \textbf{126},
230404 (2021).

\bibitem{Cabrera2018} C. R. Cabrera, L. Tanzi, J. Sanz, B. Naylor, P.
Thomas, P. Cheiney, and L. Tarruell, Quantum liquid droplets in a mixture of
Bose-Einstein condensates Science \textbf{359}, 301 (2018).

\bibitem{Cheiney} P. Cheiney, C. R. Cabrera, J. Sanz, B. Naylor, L. Tanzi,
and L. Tarruell, Bright Soliton to quantum droplet transition in a mixture
of Bose-Einstein condensates, Phys. Rev. Lett. \textbf{120}, 135301 (2018).

\bibitem{Inguscio} G. Semeghini, G. Ferioli, L. Masi, C. Mazzinghi, L.
Wolswijk, F. Minardi, M. Modugno, G. Modugno, M. Inguscio, and M. Fattori,
Self-bound quantum droplets of atomic mixtures in free space?, Phys. Rev.
Lett. \textbf{120}, 235301 (2018).

\bibitem{Inguscio2} G. Ferioli, G. Semeghini, L. Masi, G. Giusti, G.
Modugno, M. Inguscio, A. Gallem\'{\i}, A. Recati, and M. Fattori, Collisions
of self-bound quantum droplets, Phys. Rev. Lett. \textbf{122}, 090401 (2019).

\bibitem{D'Errico} C. D'Errico, A. Burchianti, M. Prevedelli, L. Salasnich,
F. Ancilotto, M. Modugno, F. Minardi, and C. Fort, Observation of quantum
droplets in a heteronuclear bosonic mixture, Phys. Rev. Res. \textbf{1},
033155 (2019).

\bibitem{Edler} D. Edler, C. Mishra, F. W\"{a}chtler, R. Nath, S. Sinha, and
L. Santos, Quantum fluctuations in quasi-one-dimensional dipolar
Bose-Einstein condensates, Phys. Rev. Lett. \textbf{119}, 050403 (2017).

\bibitem{Baillie} D. Baillie, R. M. Wilson, and P. B. Blakie, Collective
excitations of self-bound droplets of a dipolar quantum fluid, Phys. Rev.
Lett. \textbf{119}, 255302 (2017).

\bibitem{Kadau:2016} H. Kadau, M. Schmitt, M. Wenzel, C. Wink, T. Maier, I.
Ferrier-Barbut, and T. Pfau, Observing the Rosensweig instability of a
quantum ferrofluid, Nature (London) \textbf{530}, 194 (2016).

\bibitem{Ferrier} I. Ferrier-Barbut, H. Kadau, M. Schmitt, M. Wenzel, and T.
Pfau, Observation of quantum droplets in a strongly dipolar Bose gas, Phys.
Rev. Lett. \textbf{116}, 215301 (2016).

\bibitem{Chomaz:2016} L. Chomaz, S. Baier, D. Petter, M. J. Mark, F. W\"{a} 
chtler, L. Santos, and F. Ferlaino, Quantum-fluctuation-driven crossover
from a dilute Bose-Einstein condensate to a macrodroplet in a dipolar
quantum fluid, Phys. Rev. X \textbf{6}, 041039 (2016).

\bibitem{Tanzi:2019} L. Tanzi, E. Lucioni, F. Fama, J. Catani, A. Fioretti,
C. Gabbanini, R. N. Bisset, L. Santos, and G. Modugno, Observation of a
dipolar quantum gas with metastable supersolid properties, Phys. Rev. Lett.
\textbf{122}, 130405 (2019).

\bibitem{Bottcher:2019} F. B\"{o}ttcher, J.-N. Schmidt, M. Wenzel, J.
Hertkorn, M. Guo, T. Langen, and T. Pfau, Transient supersolid properties in
an array of dipolar quantum droplets, Phys. Rev. X \textbf{9}, 011051 (2019).

\bibitem{Chomaz:2019} L. Chomaz, D. Petter, P. Ilzhofer, G. Natale, A.
Trautmann, C. Politi, G. Durastante, R. M. W. van Bijnen, A. Patscheider, M.
Sohmen, M. J. Mark, and F. Ferlaino, Long-lived and transient supersolid
behaviors in dipolar quantum gases Phys. Rev. X \textbf{9}, 021012 (2019).

\bibitem{Frontiers} Z.-H. Luo, W. Pang, B. Liu, Y. Li, and B. A. Malomed, A
new form of liquid matter: quantum droplets, Front. Phys. \textbf{16}, 32501
(2021).

\bibitem{Guo:2019} M. Guo, F. B\"{o}ttcher, J. Hertkorn, J.-N. Schmidt, M.
Wenzel, H. P. Buchler, T. Langen, and T. Pfau, The low-energy Goldstone mode
in a trapped dipolar supersolid, Nature (London) \textbf{574}, 386 (2019).

\bibitem{Sabari:2017} S. Sabari, K. Porsezian and P. Muruganandam, Dynamical
stabilization of two- dimensional trapless Bose-Einstein condensates by
three-body interaction and quantum fluctuations, Chaos, Solitons and
Fractals \textbf{103}, 237 (2017).

\bibitem{Tamil:2019} R. Tamilthiruvalluvar, S. Sabari, K. Porsezian and P.
Muruganandam, Vortex formation and vortex lattices in a Bose-Einstein
condensate with Lee-Huang-Yang (LHY) correction, Physica E \textbf{107}, 54
(2019).

\bibitem{Tanzi:2019b} L. Tanzi, S. Roccuzzo, E. Lucioni, F. Fama, A.
Fioretti, C. Gabbanini, G. Modugno, A. Recati, and S. Stringari, Supersolid
symmetry breaking from compressional oscillations in a dipolar quantum gas,
Nature (London) \textbf{574}, 382 (2019).

\bibitem{Bottcher:2021} F. B\"ottcher, J.-N. Schmidt, J. Hertkorn, K. S. H.
Ng, S. D. Graham, M. Guo, T. Langen, and T. Pfau, New states of matter with
fine-tuned interactions: quantum droplets and dipolar supersolids, Rep.
Prog. Phys. \textbf{84}, 012403 (2021).

\bibitem{Hertkorn:2021} J. Hertkorn, J.-N. Schmidt, F. B\"{o}ttcher, M. Guo,
M. Schmidt, K. S. H. Ng, S. D. Graham, H. P. Buchler, T. Langen, M.
Zwierlein, and T. Pfau, Density fluctuations across the
superfluid-supersolid phase transition in a dipolar quantum gas, Phys. Rev.
X \textbf{11}, 011037 (2021).

\bibitem{Smith:2011} R. P. Smith, R. L. D. Campbell, N. Tammuz, and Z.
Hadzibabic, Effects of interactions on the critical temperature of a trapped
Bose gas, Phys. Rev. Lett. \textbf{106}, 250403 (2011).

\bibitem{Lopes:2017} R. Lopes, C. Eigen, N. Navon, D. Cl\'{e}ment, R. P.
Smith, and Z. Hadzibabic, Quantum depletion of a homogeneous Bose-Einstein
condensate, Phys. Rev. Lett. \textbf{119}, 190404 (2017).

\bibitem{Lopes:2017b} R. Lopes, C. Eigen, A. Barker, K. G. H. Viebahn, M.
Robert-de-Saint-Vincent, N. Navon, Z. Hadzibabic, and R. P. Smith,
Quasiparticle energy in a strongly interacting homogeneous Bose-Einstein
condensate, Phys. Rev. Lett. \textbf{118}, 210401 (2017).

\bibitem{Navon:2011} N. Navon, S. Piatecki, K. G\"{u}nter, B. Rem, T. C.
Nguyen, F. Chevy, W. Krauth, and C. Salomon, Dynamics and thermodynamics of
the low-temperature strongly interacting Bose gas, Phys. Rev. Lett. \textbf{ 
107}, 135301 (2011).

\bibitem{Wild:2012} R. J. Wild, P. Makotyn, J. M. Pino, E. A. Cornell, and
D. S. Jin, Measurements of Tan's contact in an atomic Bose-Einstein
condensate, Phys. Rev. Lett. \textbf{108}, 145305 (2012).

\bibitem{1999Butts} D. A. Butts and D. S. Rokhsar, Predicted signatures of
rotating Bose-Einstein condensates, Nature \textbf{397}, 327 (1999).

\bibitem{1991Donnelly} R. J. Donnelly, \textit{Quantized vortices in Helium
II} (Cambridge University Press, Cambridge, 1991).

\bibitem{Yi2006} S. Yi, and H. Pu, Vortex structures in dipolar condensates,
Phys. Rev. A \textbf{73}, 061602(R) (2006).

\bibitem{Vardi2} I. Tikhonenkov, B. A. Malomed, and A. Vardi, Vortex
solitons in dipolar Bose-Einstein condensates Phys. Rev. A \textbf{78},
043614 (2008).

\bibitem{Baranov2008} M.A. Baranov, Theoretical progress in many-body
physics with ultracold dipolar gases, Phys. Rep. \textbf{464}, 71 (2008).

\bibitem{Klawunn2008} M. Klawunn, R. Nath, P. Pedri, and L. Santos,
Transverse Instability of straight vortex lines in dipolar Bose-Einstein
condensates, Phys. Rev. Lett. \textbf{100}, 240403 (2008).

\bibitem{Klawunn} M. Klawunn and L. Santos, Phase transition from straight
into twisted vortex lines in dipolar Bose\^{a}\euro ``Einstein condensates,
New J. Phys.\textbf{11}, 055012 (2009).

\bibitem{Wilson2009} R. M. Wilson, S. Ronen, and J. L. Bohn, Stability and
excitations of a dipolar Bose-Einstein condensate with a vortex, Phys. Rev.
A \textbf{79}, 013621 (2009).

\bibitem{Abad2009} M. Abad, M. Guilleumas, R. Mayol, M. Pi, and D. M. Jezek,
Vortices in Bose-Einstein condensates with dominant dipolar interactions,
Phys. Rev. A \textbf{79}, 063622 (2009).

\bibitem{Abad2010} M. Abad, M. Guilleumas, R. Mayol, M. Pi, and D. M. Jezek,
Dipolar condensates confined in a toroidal trap: Ground state and vortices,
Phys. Rev. A \textbf{81}, 043619 (2010).

\bibitem{Malet2011} F. Malet, T. Kristensen, S.M. Reimann, and G. M.
Kavoulakis, Rotational properties of dipolar Bose-Einstein condensates
confined in anisotropic harmonic potentials, Phys. Rev. A \textbf{83},
033628 (2011).

\bibitem{Kishor2012} R. K. Kumar, and P. Muruganandam, J. Phys. B At. Mol.
Opt. Phys. \textbf{45}, 215301 (2012).

\bibitem{Wilson2012} R. M. Wilson, C. Ticknor, J. L. Bohn, and E.
Timmermans, Roton immiscibility in a two-component dipolar Bose gas, {Phys.
Rev. A} \textbf{86}, 033606 (2012).

\bibitem{Sabari:2018} S. Sabari and R. Kishor Kumar, { Effect of an
oscillating Gaussian obstacle in a Dipolar Bose-Einstein condensate}, Eur.
Phys. J. D \textbf{72}, 48 (2018).

\bibitem{Sabari2024}S. Sabari, R. K. Kumar, L. Tomio, { Vortex dynamics and turbulence in 
dipolar Bose-Einstein condensates}, 
Phys. Rev. A {\bf 109}, 023313 (2024).

\bibitem{Lauro2024}L. Tomio, A. N. da Silva, S. Sabari, R. K. Kumar,  
{Dynamical Vortex Production and Quantum Turbulence in Perturbed Bose-Einstein Condensates}, 
Few-Body Systems {\bf 65}, 13 (2024).

\bibitem{Sabari2017}S. Sabari, {Vortex formation and hidden vortices in dipolar Bose-Einstein condensates}, 
Phys. Lett. A {\bf 381}, 3062 (2017).

\bibitem{Raymond1} Y. Li, J. Liu, W. Pang, and B. A. Malomed, Matter-wave
solitons supported by field-induced dipole-dipole repulsion with a spatially
modulated strength, Phys. Rev. A \textbf{88}, 053630 (2013).

\bibitem{Raymond2} X. Jiang, Z. Fan, Z. Chen, W. Pang, Y. Li, and B. A.
Malomed, Two-dimensional solitons in dipolar Bose-Einstein condensates with
spin-orbit coupling, Phys. Rev. A \textbf{93}, 023633 (2016).

\bibitem{Raymond3} B. Liao, S. Li, C. Huang, Z. Luo, W. Pang, H. Tan, B. A.
Malomed, and Y. Li, Anisotropic semi-vortices in dipolar spinor condensates
controlled by Zeeman splitting, Phys. Rev. A \textbf{96}, 043613 (2017).


\bibitem{Sabari2025} S. Sabari, R. Sasireka, R. Radha, A. Uthayakumar, L.
Tomio, Vortices in Tunable Dipolar Bose-Einstein condensates with Attractive
Interactions, Phys. Rev. A \textbf{111}, 053320 (2025).

\bibitem{2009Bijnen} R. M. W. van Bijnen, A. J. Dow, D. H. J. O'Dell, N. G.
Parker, and A. M. Martin, Exact solutions and stability of rotating dipolar
Bose-Einstein condensates in the Thomas-Fermi limit, Phys. Rev. A \textbf{80}
, 033617 (2009).

\bibitem{Vardi} I. Tikhonenkov, B. A. Malomed, and A. Vardi, Anisotropic
solitons in dipolar Bose-Einstein condensates, Phys. Rev. Lett. \textbf{100} 
, 090406 (2008).

\bibitem{2D-LHY-vort} Y. Li, Z. Chen, Z. Luo, C. Huang, H. Tan, W. Pang, and
B. A. Malomed, Two-dimensional vortex quantum droplets, Phys. Rev. A \textbf{ 
\ 98}, 063602 (2018).

\bibitem{Leticia} Y. V. Kartashov, B. A. Malomed, L. Tarruell, and L.
Torner, Three-dimensional droplets of swirling superfluids, Phys. Rev. A
\textbf{98}, 013612 (2018).

\bibitem{Macri} A. Cidrim, F. E. A. dos Santos, E. A. L. Henn, and T. Macr 
\`{\i}, Vortices in self-bound dipolar droplets, Phys. Rev. A \textbf{98},
023618 (2018).

\bibitem{Klaus2022} L. Klaus, T. Bland, E. Poli, C. Politi, G. Lamporesi, E.
Casotti, R. N. Bisset, M. J. Mark, and F. Ferlaino, {Observation of vortices
and vortex stripes in a dipolar condensate,} Nature Phys. \textbf{18}, 1453
(2022).

\bibitem{2023Bland} T. Bland, G. Lamporesi, M. J. Mark, and F. Ferlaino, { 
Vortices in dipolar Bose-Einstein condensates,} Comptes Rendus Physique
\textbf{24}, 133 (2023).

\bibitem{Lima2011} A. R. P. Lima and A. Pelster, Quantum fluctuations in
dipolar Bose gases, Phys. Rev. A \textbf{84}, 041604(R) (2011).

\bibitem{Blakie2016} R. N. Bisset, R. M. Wilson, D. Baillie, and P. B.
Blakie, Ground-state phase diagram of a dipolar condensate with quantum
fluctuations, Phys. Rev. A \textbf{94}, 033619 (2016).

\bibitem{Wachtler2016} F. Wachtler and L. Santos, Quantum filaments in
dipolar Bose-Einstein condensates, Phys. Rev. A \textbf{93}, 061603 (2016).

\bibitem{magic} A. E. Bennett, C. M. Rienstra, M. Auger, K. V. Lakshmi, and
R. G. Griffin, Heteronuclear decoupling in rotating solids, J. Chem. Phys.
\textbf{103}, 6951 (1995).

\bibitem{2017Kumar} R. K. Kumar, L. Tomio, B. A. Malomed, and A. Gammal,
Vortex lattices in binary Bose-Einstein condensates with dipole-dipole
interactions, Phys. Rev. A \textbf{96}, 063624 (2017).

\bibitem{2012Muruganandam} P. Muruganandam and S. K. Adhikari, Numerical and
variational solutions of the dipolar Gross-Pitaevskii equation in reduced
dimensions, Laser Physics \textbf{22}, 813 (2012).

\bibitem{2017jpco} R. K. Kumar, P. Muruganandam, L. Tomio, and A. Gammal,
Miscibility in coupled dipolar and non-dipolar Bose-Einstein condensates, { 
J. Phys. Commun.} \textbf{1}, 035012 (2017).

\bibitem{2020Brito} L. Brito, A. Andriati, L. Tomio, and A. Gammal, Breakup
of rotating asymmetric quadratic-quartic trapped condensates, Phys. Rev. A
\textbf{102}, 063330 (2020).

\bibitem{Viskol} E. Shamriz, Z. Chen, and B. A. Malomed, Suppression of the
quasi-two-dimensional quantum collapse in the attraction field by the
Lee-Huang-Yang effect, Phys. Rev. A \textbf{101}, 063628 (2020).

\bibitem{2006Schutzhold} R. Schützhold, M. Uhlmann, Y. Xu, and U. R. Fischer, 
Mean-field expansion in Bose-Einstein condensates with finite-range
interactions, Int. J. Mod. Phys. B {\bf 20}, 3555 (2006).

\bibitem{2016Schmitt} M. Schmitt, M. Wenzel, F. B\"{o}ttcher, I.
Ferrier-Barbut and T. Pfau, Self-bound droplets of a dilute magnetic quantum
liquid, Nature (London) \textbf{539}, 259 (2016).

\bibitem{2001Fetter} A. L. Fetter and A. A. Svidzinsky, Vortices in a
trapped dilute Bose-Einstein condensate, J. Phys.: Condens. Matter 
\textbf{13} R135 (2001).

\bibitem{2016Kumar} R. K. Kumar, T. Sriraman, H. Fabrelli, P. Muruganandam,
and A. Gammal, Three-dimensional vortex structures in a rotating dipolar 
Bose-Einstein condensate, J. Phys. B: At. Mol. Opt. Phys. \textbf{49}, 155301 
(2016).

\bibitem{1996Baym} G. Baym and C. J. Pethick, Ground-state properties of
magnetically trapped Bose-condensed rubidium gas, Phys. Rev. Lett. \textbf{76 
}, 5 (1996).

\bibitem{1996Dalfovo} F. Dalfovo and S. Stringari, Bosons in anisotropic
traps: ground state and vortices, Phys. Rev. A \textbf{53}, 2477 (1996).

\bibitem{2016Pitaevskii} L. Pitaevskii and S. Stringari,
Bose–Einstein Condensation and Superfluidity,
Oxford University Press, Oxford, 2016 (see page 472).
\end{thebibliography}
\end{document}